\documentclass[11pt,preprint]{aastex}
\usepackage{epsfig}
\usepackage{graphicx}
\usepackage{url}
\usepackage{natbib}
\usepackage{breqn}
\usepackage{epstopdf}
\providecommand{\e}[1]{\ensuremath{\times 10^{#1}}}

\shorttitle{Testing the Dark Matter Caustic Theory Against Observations in the Milky Way}
\shortauthors{Dumas et al.}

\begin{document}

\title{Testing the Dark Matter Caustic Theory Against Observations in the Milky Way}

\author{
Julie Dumas\altaffilmark{1},
Heidi J. Newberg\altaffilmark{1},
Bethany Niedzielski\altaffilmark{1},
Adam Susser\altaffilmark{1},
Jeffery M. Thompson\altaffilmark{1},
Jake Weiss\altaffilmark{1},
Kim M. Lewis\altaffilmark{1}
}

\altaffiltext{1}{Department of Physics, Applied Physics and Astronomy,
Rensselaer Polytechnic Institute, Troy, NY 12180, USA}

\begin{abstract}
We test a particular theory of dark matter in which dark matter axions form ring ``caustics" in the plane of the Milky Way against actual observations of Milky Way stars. According to this theory, cold, collisionless dark matter particles with angular momentum flow in and out of the Milky Way on sheets. These flows form caustic rings (at the positions of the rings, the density of the flow is formally infinite) at the locations of closest approach to the Galactic center. We show that the caustic ring dark matter theory reproduces a roughly logarithmic halo, with large perturbations near the rings. We show that the theory can reasonably match the observed rotation curve of the Milky Way. We explore the effects of the caustic rings on dwarf galaxy tidal disruption using N-body simulations. In particular, simulations of the Sagittarius dwarf galaxy tidal disruption in a caustic ring halo potential match observations of the trailing tidal tail as far as 90 kpc from the Galactic center; they do not, however, match the leading tidal tail. None of the caustic ring, NFW, or triaxial logarithmic halos fit all of the data. The source code for calculating the acceleration due to a caustic ring halo has been made publicly available in the NEMO Stellar Dynamics Toolbox and the Milkyway@home client repository.
\end{abstract}

	
\section{Introduction}

\subsection{The caustic ring dark matter halo}

The caustic ring dark matter theory has been developed by Pierre Sikivie \citep{1995PhRvL..75.2911S, 2003PhLB..567....1S, 2007PhRvD..76b3505N, 2008PhRvD..78f3508D, 2011PhLB..695...22S, 2013PhRvD..88l3517B} over the past 20 years. Although Sikivie started from the spherically symmetric self-similar models of \citet{1984ApJ...281....1F} and \citet{1985ApJS...58...39B}, it is important to note that the Sikivie model has evolved to include physics of dark matter particles beyond a simple gravitational interaction. This physics results in dark matter flows that rotate with the stellar disk. 

The caustic ring model \citep{2008PhRvD..78f3508D} is consistent with a Universe that expands with the Hubble flow, and has power law density fluctuations which create the seeds of individual galaxies. The dark matter starts out on these density fluctuations with temperature zero, so that the velocity dispersion at any point in space is zero. As the dark matter falls into a Galaxy it gains velocity, but since it initially occupies only three dimensions of six-dimensional phase space (three spatial dimensions and three velocity dimensions), it falls in on discrete flows that occupy only three dimensions of the six-dimensional phase space. The infalling flows, like the particles of which they are composed, fall in toward the galaxy center until they reach their closest approach, then fly out until all of the kinetic energy has been converted to gravitational potential energy, and then turn around and fall back in.

For special locations in the Galaxy, the dark matter will form caustics, where the density is infinite in an infinitely small region of space (for non-zero velocity dispersion, the caustic will not be as sharply defined). These locations are the turnaround points on the orbits of the infalling sheets of dark matter. The caustics are rings at the inner turnaround radius, and spheres at the outer turnaround radius. In a self-similar model \citep{2011PhLB..695...22S}, the position of the caustics move outward with time, as initially more distant material collapses into the potential well of the galaxy. The number of caustics increases as more matter falls into the galaxy (with infall times of about $10^8$ years); each time the initial infalling particles reach perigalacticon, a new ring caustic forms near the center of the galaxy, at the original Galactocentric distance of that earliest accreting matter. Thus, the number of caustics grows with time (about one additional caustic for each $10^8$ years); the most recently accreted matter is associated with the outermost ring. The locations of the ring caustics depend only on the slope of the evolved power spectrum of density perturbations on galaxy scales, the rotation speed of the galaxy, and the age of the Universe. For the Milky Way (rotation speed of 220 $\rm km\: s^{-1}$ at 8.5 kpc from the Galactic center) and the current standard cold dark matter ($\Lambda$CDM) model of the Universe (age = 13.7 Gyr), the inner ring caustics are predicted to be at radius:
\begin{equation}
a_n\simeq\frac{40 \: {\rm kpc}}{n} \; \; \; (n=1, 2, 3, 4, ...)
\end{equation}
The first four caustics in the Milky Way are rings at predicted distances of 40 kpc, 20 kpc, 13 kpc, and 10 kpc from the center of the Milky Way \citep{2008PhRvD..78f3508D}. In cross section, these rings have a tricusp shape, as shown in Figure \ref{g-ring} (top panel).

\subsection{Caustic rings and hierarchical structure formation}
Although the caustic ring model (we will also refer to it as the caustic ring halo) does not include the effects of hierarchical merging, it is plausibly consistent with the merger history of the Milky Way. Large galaxy mergers will destroy dark matter caustics similarly to the way galaxy mergers disrupt disks of stars \citep{1992ApJ...389....5T}, and any other coherent structure of collisionless particles. However, it is very important to note that any dark matter that falls in after the collision will remain intact on the 3D sheets in 6D phase space. A merger produces a local thickening in the phase space sheet, and a velocity dispersion associated with the flow of dark matter in a caustic ring. \citet{2003PhLB..567....1S} estimates the spread of the caustic ring radius, $\delta a$, due to the velocity dispersion, $\delta v$, of the flow of dark matter. Minor mergers will cause the caustics to spread and become fuzzy, but the caustic will become sharp again after the merger event, as material from the cold flow fills the caustic in again. Minor mergers like the Sagittarius dwarf galaxy \citep{1994Natur.370..194I} that is currently being disrupted by the Milky Way's tidal forces would not affect the structure of the caustics.

The relationship between the redshift ($z$) at which the caustic ring forms and ring number ($n$) is approximately (P. Sikivie, private communication):
\begin{equation}
z(n) = (3.8 n - 0.9)^{0.4}-1.
\end{equation}
Therefore the last ($n=1$ is the last ring to form) fourteen caustic rings ($n = 1, 2, .... 14$) were formed after $z = 3.9$, and the last two caustic rings ($n = 1, 2$) were formed after $z = 1.1$. If there were no major mergers since $z = 1.1$, then we would expect to see a ring caustic ($n = 2$) at 20 kpc from the Galactic center, just where we see the Monoceros Ring of stars. 

In the Milky Way in particular, there is no evidence for a major merger in recent history. In particular, the boxy bulge of the Milky Way is consistent with formation from disk instabilities rather than from mergers \citep{2013MNRAS.432.2092N, 2010ApJ...720L..72S}. Even within the $\Lambda$CDM model, galaxies with the lowest ratios of bulge to total luminosity have no major mergers at $z<2$ \citep{2012ApJ...756...26M}. \citet{2002ApJ...574L..39G} and \citet{2006ApJ...639L..13W} have suggested that the thick disk may be the result of the last major merger of the Milky Way, $10-12$ Gyr ago ($1.8< z < 4$). If this is true, the four ring caustics, at 40 kpc, 20 kpc, 13 kpc, and 10 kpc from the center of the Milky Way, should be undisturbed.

The details of the caustic ring model are still evolving; some of the effects of the Bose-Einstein condensate (BEC) model are still being incorporated in the model \citep{2013PhRvD..88l3517B}. The axions in this special degenerate condensed state (the de Broglie wavelength, $\hbar/p$, must be large compared to inter-particle dimensions) set up a vorticity that is more permanent than the original model, when the axion self-interactions are included (the particles share angular momentum among each other). Just as water going down the drain of a sink has a curl (quasi-rigid rotation with the inner particles rotating faster than the outer particles), axions in a BEC also have a curl. In contrast, the velocity field for WIMPs has zero curl. Although the original density fluctuations are spherical by construction, the vorticity of the axions will result in an oblate dark matter halo. In addition, the baryons and the BEC dark matter may thermalize, thus explaining why disks have such a high angular momentum. Although very recent $\Lambda$CDM cosmological simulations can reproduce disks that are plausibly like those observed, it is unclear what process gives the disks rotation, while leaving the dark matter halo without significant rotation \citep[see][]{2014MNRAS.437.1750M}. In this paper, we do not include the modifications to the caustic ring model suggested by \citet{2013PhRvD..88l3517B} because their implications on the Galactic potential have not yet been worked out.

\subsection{Implications of a caustic ring halo on the nature of dark matter}

The density fluctuations that are the precursors of galaxies in the standard caustic ring halo are by construction spherical. If the dark matter is additionally without rotation, then the dark matter falls in along radial orbits, and the inner caustic forms a cusp at the center of each galaxy, with a density that diverges as $1/r^2$ at the center \citep{1995PhRvL..75.2911S}. It is only when the dark matter has net rotation that the caustics form tricusp rings in the plane perpendicular to the axis of rotation, at the positions of the inner turnaround radii of the dark matter flows.  This is in sharp contrast to $\Lambda$CDM simulations, in which the angular momentum of the dark matter distribution in most simulated galaxies is small. It requires a special form of dark matter to form galaxies with sufficient angular momentum to form the proposed caustic ring halos.  

\citet{2011PhLB..695...22S} showed that the dark matter distribution in galaxies has sufficient angular momentum if the dark matter is composed of axions rather than WIMPs or sterile neutrinos. The difference arises from the fact that the axions form a Bose-Einstein condensate (BEC) in which the particles exchange angular momentum through gravity. In general, the angular momentum of a galaxy as a whole is thought to be caused by a tidal torque from nearby protogalaxies early in its formation, before the galaxy has condensed out of the expansion of the Universe (\citealt{1953ApJ...118..513H}; \citealt{1969ApJ...155..393P}; \citealt{1971A&A....11..377P}; \citealt{1980ComAp...8..169E} gives a review), though the dark matter can pick up net rotation through hierarchical merging \citep{2002ApJ...581..799V, 2011MNRAS.415.2607D}. Since dark matter is typically modeled as collisionless particles, one would not normally expect that the dark matter would be susceptible to a tidal torque. However, axions in a rethermalizing Bose-Einstein condensate will react to a tidal torque by going to the lowest energy state compatible with the acquired angular momentum. That state is a state of rigid rotation on the turnaround sphere. The surprising suggestion that axions would form a Bose-Einstein condensate was confirmed by \citet{2013PhRvD..87h5010S}. If solid evidence of large angular momentum in the dark matter component of the galaxy is obtained, that would be an argument in favor of axion dark matter, or at least for additional dark matter physics.

\subsection{Evidence for a caustic ring dark matter halo}

Some observational evidence that caustic rings might exist in galaxies has been put forward by previous authors. This evidence includes: the statistical distribution of bumps in the rotation curves of 32 galaxies (\citealt{2000PhRvD..61h7305K}, using data from \citealt{1991MNRAS.249..523B}, and from \citealt{1996ApJ...473..117S}); the distribution of bumps in the rotation curve of the Milky Way (\citealt{2003PhLB..567....1S}, using data from \citealt{2000MNRAS.311..361O}); a triangular feature in the Infrared Astronomical Satellite (IRAS; http://skyview.gsfc.nasa.gov) map in a direction tangent to the nearest ($n=5$) caustic ring, and aligned with the expected position of the tricusp caustic \citep{2003PhLB..567....1S}; and the Monoceros Ring at the position of the $n=2$ caustic ring (\citealt{2007PhRvD..76b3505N}, citing \citealt{2002ApJ...569..245N}). In order to match the rotation curves of the 32 external galaxies with the caustic ring model (which has only been worked out for the Milky Way rotation speed), the rotation curves were rescaled by (220 $\rm km\: s^{-1}$) $v^{-1}$, where $v$ is the actual rotation speed of the galaxy. After this rescaling, there were 3 sigma detections of discrepant velocities in the rotation curve at (rescaled) radii of about 20 and 40 kpc from the galaxy centers. The need for rings of dark matter to explain the inner part of the Milky Way galaxy rotation curve was also noted by \citet{2009PASJ...61..227S} and \citet{2011JCAP...04..002D}, who proposed two rings of dark matter at distances of 4.2 and 12.4 kpc from the Galactic center to explain dips in the rotation curve. The rings are close to Sikivie's prediction of an $n=3$ caustic ring at 13.6 kpc and an $n=10$ caustic ring at 4.3 kpc from the Galactic center, although it seems reasonable that the $n>10$ rings might not be visible in the Milky Way rotation curve (because the distance between rings is less than a kpc or they were destroyed by ancient mergers). Possibly, the caustic ring model could be improved; for instance it does not currently include the presence of a bar in the disk or the effects of hierarchical merging. It is also interesting to note that \citet{2003PhLB..567....1S}, using data from the Massachusetts-Stony Brook North Galactic Plane CO survey \citep{1985ApJ...295..422C, 1986ApJS...60....1S, 1986ApJS...60..297C}, found evidence for about ten ripples in the rotation curve between 3 and 8.5 kpc from the Galactic center, which is about what was expected from caustics. The triangular feature in the IRAS maps, if caused by a tricusp caustic, implies $p \sim 130$ pc and $q \sim 200$ pc, where $p$ and $q$ are the horizontal and vertical sizes of the caustic.

\section{Calculating the Caustic Ring Gravitational Field}

Note that there is no way to add the caustic rings to a standard halo density model such as an NFW profile \citep{1996ApJ...462..563N}, because the acceleration calculated in the caustic ring model is not the acceleration due to an excess of matter on a ring; it is rather the acceleration due to one sheet of dark matter falling in to the Milky Way. To derive the gravitational field for the entire halo, one must perform a vector sum of all of the accelerations due to all of the 20 $(n=1, 2, 3, 4, ...)$ sheets of matter that have fallen in within the last $2\e{9}$ Gyr (previous infalls are expected to be a small fraction of the total halo mass). All of the halo dark matter, including the ring caustics, is included in this final calculation of the acceleration.  There is no freedom to adjust the overall shape of the halo; it simply falls out of the sum of the integrals.

\begin{figure}
\epsscale{0.67}
\plotone{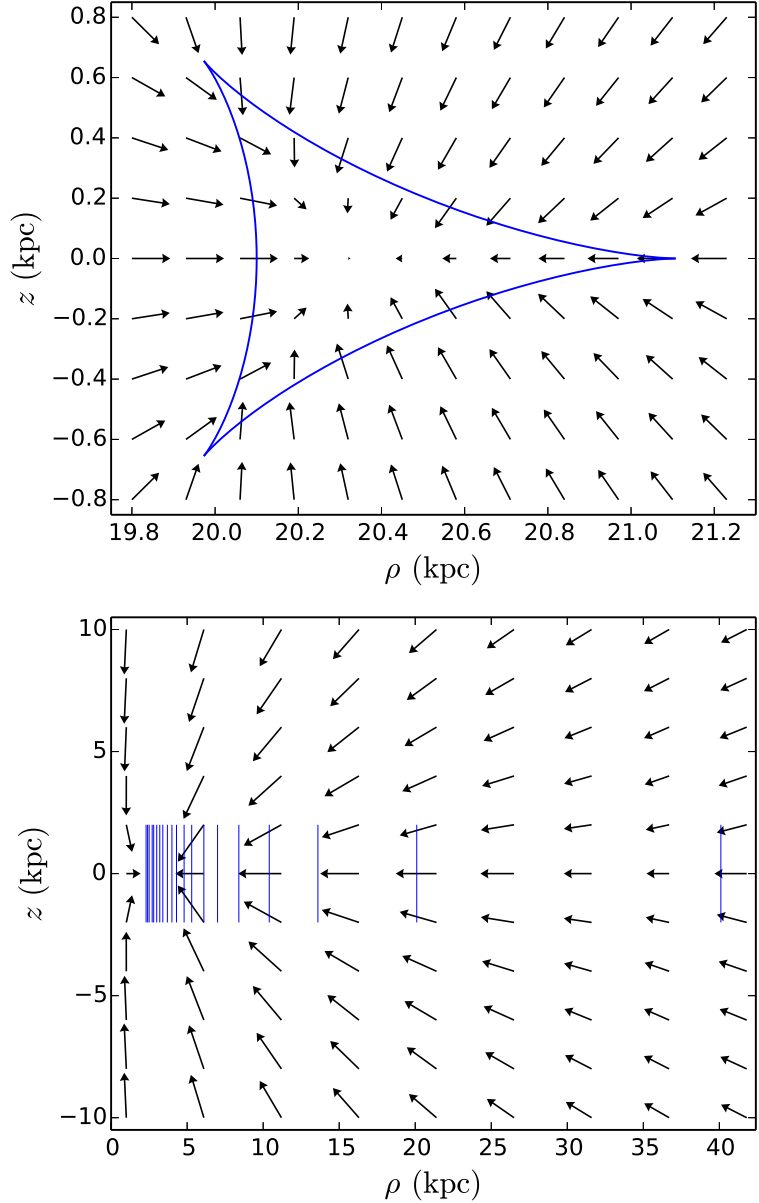}
\caption{(Top) Vector representation of gravitational field $\vec{g}_{\rm near}$ for the flow of dark matter near the $n=2$ caustic ring. $\rho$ and $z$ are Galactocentric cylindrical coordinates. The blue solid line outlines the `tricusp' shaped ring. (Bottom) Vector representation of $\vec{g}_{\rm halo}$ for the sum of the flows from all caustic rings ($n=1-20$) in the Milky Way. The vertical blue lines are at the postions of the caustic rings. $\rho$ and $z$ are Galactocentric cylindrical coordinates.}
\label{g-ring}
\end{figure}

Since the caustic ring model is axially symmetric, the equations for dark matter density and gravitational field in the Galaxy are conveniently expressed in cylindrical coordinates $(\rho,z)$, centered on the Galactic center. There is one equation for the density at a position near a caustic ring and another for the density at a position far away from a caustic ring. The equation for the density near a caustic is given by Equation 2 in \citet{2007PhRvD..76b3505N}:
\begin{equation}
d(\rho,z)_{\rm near}=\frac{1}{\rho} \sum\limits_{j=1}^{N(\rho,z)} \frac{dM}{d\Omega dt}(\alpha,\tau)   \left. \frac{\cos\alpha}{|D(\alpha,\tau)|} \right|_{(\alpha_j(\rho,z), \tau_j(\rho,z))}.
\label{eq-dnear}
\end{equation}
Here the sum is over the number of flows of axions at $(\rho,z)$, $\frac{dM}{d\Omega dt}(\alpha,\tau)$ is the rate at which mass ($dM$) falls in to the position $(\rho,z)$ per unit solid angle and unit time ($d\Omega dt$), and $D(\alpha,\tau)$ is the Jacobian determinant of the transformation between the coordinates $(\alpha,\tau)$ and $(\rho,z)$. Each of these quantities is a complex function of constants and coordinates in $(\alpha,\tau)$, defined in \citet{2007PhRvD..76b3505N}. $\alpha$ is an angle associated with the particle's last turnaround and $\tau$ is the time when the particle crosses the $z=0$ plane. The density of dark matter near a caustic ring is very large because there are many particles at turnaround points on their orbits around the Galaxy. The near equation shows this behavior since the density $d(\rho,z)$ approaches infinity when $D(\alpha,\tau)$ approaches 0. When $D(\alpha,\tau)=0$, the transformation between the coordinates $(\alpha,\tau)$ and $(\rho,z)$ gives a set of points $(\rho,z)$ outlined by a tricusp shape. The tricusp shape is the blue solid line in Figure \ref{g-ring} (top panel).

The equation for the density far from a caustic is given by Equation 3.45 in \citet{2008PhRvD..78f3508D}:
\begin{equation}
d(\rho,z)_{\rm far}=\frac{1}{v}\frac{dM}{d\Omega dt}\frac{1}{\sqrt{(r^2-a^2)^2+4a^2z^2}},
\end{equation}
where $r=\sqrt{\rho^2+z^2}$, $v$ is the speed of the flow near the caustic, $\frac{dM}{d\Omega dt}$ is the mass infall rate per unit solid angle and unit time, and $a$ is the caustic ring radius from the Galactic center. The equations given above for density apply to one caustic ring and its associated infalling and outgoing axions. The predicted radii $(a_n)$, flow speeds $(v_n)$, and infall rates $(\frac{d^2M}{d\Omega dt}|_n)$, for the first 20 caustics in the disk of the Galaxy are listed in Table 3 of \citet{2008PhRvD..78f3508D}. These parameters will now be used with the subscript $n$ that specifies the flow number ($n=1\; {\rm to}\; 20$).

The gravitational field for each caustic ring is similarly expressed by a near and far equation. If the curved nature is neglected near the ring, it can be modeled as a straight tube. This leads to a gravitational field of the following form:
\begin{equation}
\vec{g}(\rho,z)=-2G\int d\rho'dz'\hspace{.1cm}d(\rho',z')\frac{(\rho-\rho')\hat{\rho}+(z-z')\hat{z}}{(\rho-\rho')^2+(z-z')^2},
\label{eq-gnear}
\end{equation}
\citep{2007PhRvD..76b3505N} where $G$ is the gravitational constant and $d(\rho',z')$ is the dark matter density (Equation \ref{eq-dnear}). This equation describes the gravitational field from the flow of all axions flowing into and out of the ring, and not just the ring structure itself. \citet{2012arXiv1205.1260T} presents a method to calculate this rather difficult integral analytically. After transforming coordinates and substituting $d(\rho,z)$ from Equation \ref{eq-dnear} into Equation \ref{eq-gnear}, \citet{2012arXiv1205.1260T} reduces $\vec{g}$ into the form:
\begin{equation}
\vec{g}(\rho,z)_{\rm near}=-\frac{8 \pi G}{\rho v_n}\frac{dM}{d\Omega dt}|_n \hspace{.1cm} [I_\rho(\rho,z)\hat{\rho}+I_z(\rho,z)\hat{z}],
\label{eq-gnear2}
\end{equation}
where $I_\rho(\rho,z)$ and $I_z(\rho,z)$ are integrals defined in \citet{2012arXiv1205.1260T}. Here we replace the $b$ in Equation 16 from \citet{2012arXiv1205.1260T} with $v_n$. $b$ is of the same order of magnitude as the speed of the particles in the caustic \citep{2007PhRvD..76b3505N}. Figure \ref{g-ring} (top panel) shows the vector representation of $\vec{g}(\rho,z)_{\rm near}$ for the n=2 caustic ring which lies at the radius $a_n$ of 20.1 kpc. The field points towards the caustic in all directions. The tricusp is symmetric around a central point where the field goes to zero. For the n=2 ring (Figure \ref{g-ring}; top panel), the central point lies at $(\rho,z)=(20.35,0)$ kpc. 

The gravitational field at a position far from the caustic is:
\begin{equation}
\vec{g}(\rho,z)_{\rm far}=\frac{A_n}{s(2a_n^2+s)}[(r^2-a_n^2)\rho\hat{\rho}+(r^2+a_n^2)z\hat{z}],
\label{eq-gfar}
\end{equation}
\begin{equation}
A_n=\frac{4 \pi G}{v_n}\frac{dM}{d\Omega dt}\big|_n,
\label{eq-An}
\end{equation}
\vspace{.2cm}
where $r=\sqrt{\rho^2+z^2}$ and $s=\sqrt{(r^2-a_n^2)^2+4a_n^2z^2}$ (P. Sikivie, private communication).

The caustic ring halo acceleration vs. Galactocentric distance in Figures \ref{g-ring} (bottom panel) and \ref{accel} was calculated as follows: if the distance ($\rho$, $z$) is inside the tricusp ring, use the field near the ring (Equation \ref{eq-gnear2}); if the distance ($\rho$, $z$) is outside the ring use the field far from the ring (Equation \ref{eq-gfar}). The gravitational field at a distance ($\rho$, $z$) is the vector sum from all caustic rings ($n=1\; {\rm to}\; 20$):
\begin{equation}
\vec{g}(\rho,z)_{\rm halo}=\sum\limits_{n=1}^{20}\vec{g}_n(\rho,z).
\end{equation} 
We spent considerable effort finding the spatial limits that made the near-caustic function match seamlessly to the far-caustic function, and finding and deciphering the published model parameters for the Milky Way \citep{2008PhRvD..78f3508D}. There were essentially no tunable parameters in the model. Figure \ref{g-ring} (bottom panel) shows the gravitational field vectors for $\vec{g}_{\rm halo}$ as a function of position in the Milky Way. The blue lines show the positions of one cross section of the caustic rings in the $\rho z$ plane. The field generally points towards the Galactic center except near the innermost ring, where the field points away from the center. This is because there is no mass inside the $n=20$ ring in our model, so objects not exactly at the galaxy center are pulled towards the innermost rings. The contribution to the halo mass from the additional caustics for $n>20$ is assumed to be negligible. These additional caustics have likely been rearranged by early merger events. Although these figures show the cross section of the rings in the $\rho z$ plane, the tricusp is barely resolved on this scale. This is because the average horizontal size of the tricusp is on the order of $\sim0.4$ kpc. Note the vertical size of the tricusp has been exaggerated in Figure \ref{g-ring} (blue lines in bottom panel); this average size is $\sim0.5$ kpc. The sizes of the tricusp are not predicted by the caustic theory. \citet{2008PhRvD..78f3508D} estimate these sizes from rises in the Milky Way rotation curves, which will be discussed later in this paper. The reasoning is that (at $z=0$) there are discontinuities in the rotation curve at positions that lie on the boundary of the ring. For example, in Figure \ref{g-ring} the tricusp boundary for $z=0$ lies at $\rho=20.1$ kpc and $\rho=21.1$ kpc. This implies there will be discontinuities in the rotation curve at $\rho=20.1$ kpc and $\rho=21.1$ kpc.

The source code for calculating the caustic ring halo acceleration given a position has been made available on the Milkyway@home GitHub repository (\url{http://github.com/Milkyway-at-home/milkywayathome_client}). The caustic ring halo has also been released as a gravitiational potential module in the NEMO stellar dynamics toolbox (\url{http://bima.astro.umd.edu/nemo/}). Orbits and N-body simulations can be run on particles in a caustic ring halo potential using both NEMO and Milkyway@home.

\section{Comparison of the Caustic Ring Halo with Other Halo Models}

\begin{figure}
\epsscale{0.9}
\plotone{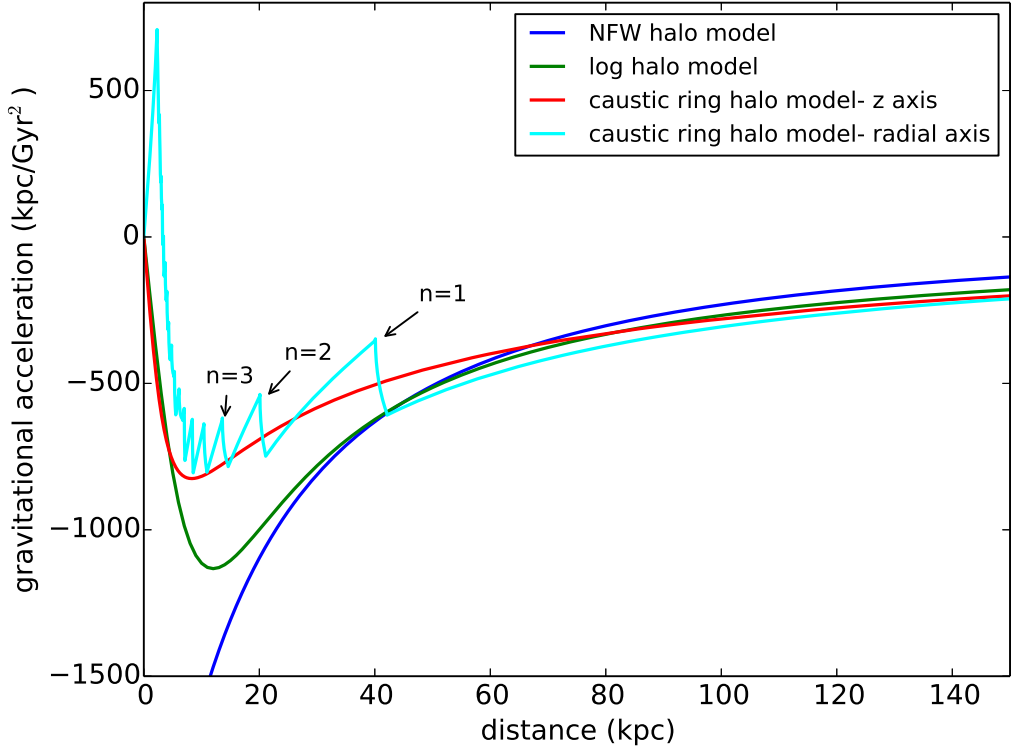}
\caption{Gravitational acceleration vs. distance from the Galactic center. The cyan line shows the acceleration in the $\rho$ direction for points in the Galactic plane ($z=0$) as a function of $\rho$. Note the bumps at positions where caustic rings are predicted. The bumps near the first three caustic rings at 40.1 ($n=1$), 20.1 ($n=2$), and 13.6 ($n=3$) kpc from the Galactic center are labeled. The red line shows the acceleration in the $z$ direction for points along the axis of symmetry ($\rho=0$) as a function of $z$. Since there are no caustics along the axis of symmetry of the Galaxy, no peaks appear in the red curve. The acceleration of the NFW (blue) and logarithmic halo (green) are shown for comparison; since these two potentials are spherically symmetric, the acceleration as a function of $\rho$ and $z$ are the same.}
\label{accel}
\end{figure}

The halo potential models (NFW and logarithmic) that will be compared to the caustic ring model throughout this paper are listed below. $X$, $Y$, and $Z$ are Galactocentric Cartesian coordinates for all models used in this paper. 

The equation for the spherical NFW model \citep{1996ApJ...462..563N} was taken from Equation 2.67 in \citet{2008gady.book.....B}:
\begin{equation}
\Phi_{\rm NFW}=-4\pi G\rho_0r_s^2\frac{\ln({1+r/r_s})}{r/r_s},
\label{nfw}
\end{equation}
where $r=\sqrt{X^2+Y^2+Z^2}$. $\rho_0$ is a characteristic overdensity of the halo and $r_s$ is the scale radius. 
The parameters we used for the NFW model were adopted from \citet{2008ApJ...684.1143X}. In this paper, the observed radial velocities of BHB stars from the Sloan Digital Sky Survey \citep[SDSS;][]{2000AJ....120.1579Y} were compared to those from simulations of the formation of galaxies like the Milky Way. In \citet{2008ApJ...684.1143X}, $\rho_0$ is defined as $\rho_s$ \citep[Equation 14 in][]{2008ApJ...684.1143X}, and $r_s=r_{vir}/c$. The characteristic density $\rho_s$ depends on the critical density of the Universe, the matter contribution to the critical density, and the critical overdensity at virialization. $r_{vir}$ is the virial radius and $c$ is a dimensionless concentration parameter. A Milky Way model including an NFW halo was fit to the rotation curve. \citet{2008ApJ...684.1143X} found fit values for the virial radius and concentration. We use those fit parameters here for the NFW model. The values for the parameters in Equation \ref{nfw} are listed in Table \ref{halomodel}. 

The parameters for a logarithmic (log) halo were taken from \citet{2005ApJ...619..807L}. \citet{2005ApJ...619..807L} used a log halo in the Galaxy model to fit the orbit of the Sagittarius dwarf galaxy tidal stream. We discuss the Sagittarius stream in Section 5. $v_{\rm halo}$ is the halo speed, $q$ is the $Z$-direction flattening, and $d$ is the halo scale length. The functional form of the logarithmic halo is: 
\begin{equation}
\Phi_{\rm log}=v_{\rm l,\,halo}^2\ln\bigg({{X^2+Y^2+}\frac{Z^2}{q_{\rm l}^2}+d^2}\bigg).
\end{equation}
The parameters for a triaxial halo were taken from \citet{2010ApJ...714..229L}. $v_{\rm halo}$, q, and $d$ are defined the same as they are for the logarithmic halo, and the functional form is:
\begin{equation}
\Phi_{\rm triaxial}=v_{\rm t,\,halo}^2\ln\bigg({C_1{X^2}+C_2{Y^2}+C_3{XY}+\frac{Z^2}{q_{\rm t}^2}+d^2}\bigg).
\end{equation}
The constants $C_1$, $C_2$, and $C_3$ are defined in Equations 4-6 of \citet{2010ApJ...714..229L}. These allow for rotation about the halo Z axis and flattening in the $X$ and $Y$ directions. These parameters were also fit by \citet{2010ApJ...714..229L} to match the orbit of the Sagittarius tidal stream and are listed in Table \ref{halomodel}.

\begin{deluxetable}{ccc}
\tablecolumns{3}
\tablewidth{0pt}
\tablecaption{Halo Model Parameters}
\tablehead{\colhead{Model} & \colhead{Parameter} & \colhead{Value}} 
\startdata
NFW & Characteristic density & $\rho_o=4.197\e{-3}\: M_{\odot}\: \rm pc^{-3}$\\
    & Scale radius & $r_s=22.25\: \rm kpc$\\
\\[-0.4cm]
\tableline
\\[-0.4cm]
logarithmic & Halo speed & $v_{\rm l,\, halo}=114.0\: \rm km\: s^{-1}$\\
            & $Z$ direction flattening & $q_{\rm l}=1.0$\\
\\[-0.4cm]
\tableline
\\[-0.4cm] 
triaxial & Halo speed & $v_{\rm t,\, halo}=121.9\: \rm km\: s^{-1}$\\
         & $Z$ direction flattening & $q_{\rm t}=1.36$\\
         & Scale length & $d=12.0\: \rm kpc$\\
         & Constant & $C_1=0.992947$\\
         & Constant & $C_2=0.532153$\\
         & Constant & $C_3=0.114889$\\
\enddata
\label{halomodel}
\end{deluxetable}

The gravitational acceleration of the NFW and spherical ($q_{\rm l}=1.0$) log halo models in Figure \ref{accel} was calculated using $\vec{g}=-\vec{\nabla}\Phi$. It is important to note that unlike the NFW, log, and triaxial models, the caustic ring model parameters are predicted directly from theory. The caustic ring model parameters $a_n$, $v_n$, and $\frac{dM}{d\Omega dt}\big|_n$ (Equations \ref{eq-gnear2}-\ref{eq-An}) follow from solutions to the equations of motion for axion flows falling into the Galaxy. The initial conditions are specified by the initial dark matter mass distribution, the rotation speed of the Galaxy, and the initial dark matter flow angular momentum. The `tunable' parameters are the angular momentum parameter $j_{max}$ and the size $p$ of the caustic ring  \citep{2008PhRvD..78f3508D}. The angular momentum parameter  $j_{max}$ for the Milky Way was fit to the rotation curve by \citet{1995PhRvL..75.2911S}. The model predicts the radii for the first 20 caustics \citep{2008PhRvD..78f3508D}. The contribution to the halo mass from the additional caustics for $n>20$ is assumed to be negligible. The sizes $p$ of the caustic rings are inferred from the analysis of bumps in rotation curves. Table 4 of \citet{2008PhRvD..78f3508D} lists the positions of observed rises in Milky Way rotation curves corresponding to the positions of the $n=3$ and $n=5-14$ caustic rings. There should be an upward kink in the rotation at the position of one edge of the caustic and a downward kink at the position of the other edge of the caustic. In Figure \ref{g-ring} (top panel), this corresponds to the left edge at $\rho=20.1$ kpc and the right edge at $\rho=$ 20.4 kpc. For our calculations of the caustic ring model gravitational acceleration we use the $p$ values from Table 4 of \citet{2008PhRvD..78f3508D} for the $n=3$ and $n=5-14$ rings. For $n=1-2,\: 4$, and $15-20$ we use $p_n/a_n=0.05$ which is the average value of the ratio $p_n/a_n$ for the other rings.

The caustic ring halo acceleration is compared to that of an NFW halo and a spherical log halo in Figure \ref{accel}. The acceleration for the NFW and log halos are given as a function of distance from the Galactic center. For the caustic ring halo, which is not spherically symmetric but is axially symmetric, we show acceleration in the plane ($z=0$) as a function of cylindrical radius $\rho$, and along the axis of symmetry ($\rho=0$) as a function of z. In either case the direction of the acceleration is towards the Galactic center. The obvious feature is that in the Galactic plane there are large jumps in the acceleration at the positions of the caustic rings. Since the NFW and log halos are spherically symmetric, there is no difference in these curves for $z=0$ and $\rho=0$. 

Given the complexity of calculating the acceleration, it is extremely gratifying and reassuring to discover that the resulting potential was plausibly similar to familiar Galactic potentials. Note that the caustic ring halo, overall, is reasonably similar to a log halo, but with jumps in the acceleration near the caustics. The  shape of the dark matter profile differs tremendously from the NFW profile near the Galactic center; there is no strong central cusp in the caustic ring halo. In fact, it is just the opposite; there is essentially {\it no} dark matter in the center of a caustic ring halo. The sharp spike in Figure \ref{accel} near the center indicates an acceleration away from the Galactic center, towards the caustic rings. Figure \ref{g-ring} (bottom panel) shows the gravitational field vectors as a function of position in the Milky Way, showing the positive acceleration along the radius inside the innermost ring caustic. This affect disappears when bulge and disk potentials are included, as we shall see later. In the real galaxy, we can imagine that the caustic ring halo is slightly blurred due to the non-zero velocity dispersion of the dark matter. Also, past major mergers would disrupt the caustic pattern near the Galactic center.

\section{Comparison of the Caustic Ring Halo with the Milky Way Rotation Curve}

The potential of the Milky Way galaxy includes contributions from the dark matter halo, the disk of baryons, and the stellar bulge. In this paper, we use analytical disk and bulge potentials to model the baryons in stars and gas in the Milky Way. We use a \citet{1975PASJ...27..533M} disk and \citet{1990ApJ...356..359H} bulge:
\begin{equation}
\Phi_{\rm disk}=-\frac{GM_{\rm disk}}{\sqrt{{\rm X^2+Y^2}+(a+\sqrt{{\rm Z^2}+b^2})^2}},
\end{equation}
\begin{equation}
\Phi_{\rm bulge}=-\frac{GM_{\rm bulge}}{r+c},
\end{equation}
where $r=\sqrt{\rm X^2+Y^2+Z^2}$. $M_{\rm disk}$ and $M_{\rm bulge}$ are the masses of the disk and bulge. $a$ and $b$ are the disk scale length and height. $c$ is the bulge scale radius. $X$, $Y$, and $Z$ are Galactocentric Cartesian coordinates for all models used in this paper. We will compare the results of three cases for the fixed analytical potential of the Milky Way:
\begin{enumerate}
\item
$\rm \Phi_{Galaxy}= \Phi_{Hernquist \; bulge}+\Phi_{Miyamoto-Nagai \; disk}+\Phi_{caustic \; halo}$
\item
$\rm \Phi_{Galaxy}= \Phi_{Hernquist \; bulge}+\Phi_{Miyamoto-Nagai \; disk}+\Phi_{logarithmic \; halo}$
\item
$\rm \Phi_{Galaxy}= \Phi_{Hernquist \; bulge}+\Phi_{Miyamoto-Nagai \; disk}+\Phi_{triaxial \; halo}$.
\end{enumerate}
The three halo models used here are discussed in Section 3. The rotation curve was obtained as follows. For circular orbits the centripetal acceleration at a distance $r$ from the Galactic center is $g(r)=\frac{v(r)^2}{r}$, the square of the velocity at at $r$ divided by $r$. We get the acceleration from from the Miyamoto-Nagai disk, Hernquist bulge, log halo, and NFW halo potential models using $\vec{g(r)}=-\vec{\nabla}\Phi$. For the caustic ring halo, we already have the acceleration. To get the contribution from the gas in the disk, we used the velocity points of the gas disk model from \citet{2000MNRAS.311..361O}. Our $g_{\rm gas}$ is a fifth order polynomial fit to the HI gas model and a sixth order polynomial fit to the $\rm H_2$ gas model. Since we have the acceleration for all of our models, we can solve for $v(r)$. We add all the contributions to the Galaxy acceleration $g_{\rm total}=g_{\rm disk}+g_{\rm bulge}+g_{\rm gas}+g_{\rm halo}$ and plot $v=\sqrt{g_{\rm total}\cdot r}$ for the rotation curve.
\begin{figure}
\epsscale{0.9}
\plotone{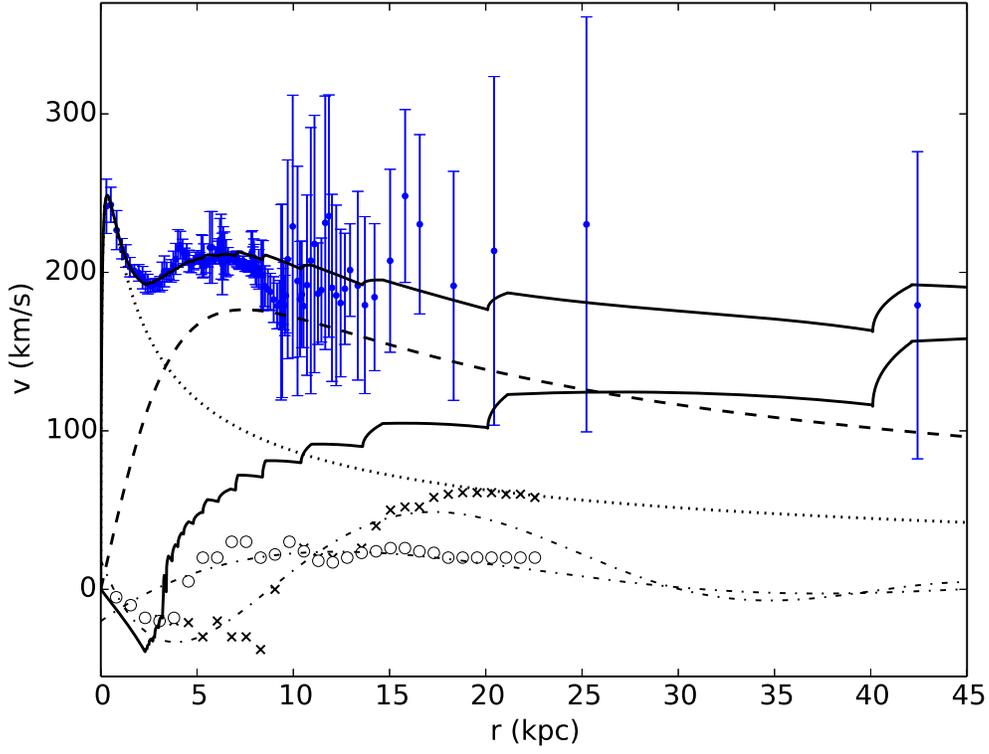}
\caption{Rotation curve of the Milky Way. The sum of a caustic ring halo (lower solid line), Hernquist bulge (dotted), Miyamoto-Nagai disk (dashed), HI gas layer (dashed-dotted line), and $\rm H_2$ gas layer (dashed-dotted line) produce a rotation curve (upper solid line) that matches observations. The blue data points are from \citet{2012PASJ...64...75S}. The dashed-dotted line is a polynomial fit to the gas model (open circles for $\rm H_2$ and crosses for HI) from \citet{2000MNRAS.311..361O}.}
\label{rc}
\end{figure}

We use a reduced $\chi^2$ as a measure of fitness of the rotation curve data points to the models: 
\begin{equation}
\chi^2_{v}=\frac{1}{N-n-1} \sum\limits_{i} \left(\frac{v_{model,i}-v_{data,i}}{\sigma_{v,i}}\right)^2,
\label{chisq1}
\end{equation}
where $v$ is the velocity, $\sigma_{v,i}$ is the error in the measurement of the data point, $N$ is the number of data points, and $n$ is the number of parameters being fit. We performed the fit by finding the optimum $\chi^2$ value for rotation curves produced with a given set of parameters for the disk, bulge, and halo models. Reduced $\chi^2$ values around 1.0 represent a reasonable fit to the data, implying most of the model point values are within the errors of the data point values. The parameters for the log \citep[parameter values adopted from][see Section 3]{2005ApJ...619..807L}, NFW \citep[parameter values adopted from][see Section 3]{2008ApJ...684.1143X}, and caustic ring halo were not fit. 

In this case, the $\chi^2$ value now depends only on the disk and bulge parameters: $\chi^2(M_{\rm disk}, \; a, \; b,$ $M_{\rm bulge}, \; c)$. The fit parameters were found by parameter sweeps; varying one parameter while leaving the others constant and finding the chi-squared values. This consequentially finds the $\chi^2$ as a function of one parameter. For example, if we have an arbitrary parameter $p$, we can find $\chi^2(p)$. If the plot of $\chi^2(p)$ is close to a horizontal line, the fit doesn't depend on this parameter. Parameters that affect the fit produce $\chi^2(p)$ plots that are close to parabolas with a dip at the minimum chi-squared value. We found that the fit here depends on the disk mass $(M_{\rm disk})$, disk scale length $(a)$, bulge mass $(M_{\rm bulge})$, and bulge scale length $(c)$ but doesn't depend on the value for the disk scale height $(b)$. We have $\chi^2(M_{\rm disk}, \; a, \; M_{\rm bulge}, \; c)$ with 4 free parameters. The number of data points is $N=121$ and the number of parameters is $n=4$ (Equation \ref{chisq1}). The fit values we found for the disk and bulge parameters are shown in Table \ref{rcfit}.

\begin{deluxetable}{lcccc}
\tablecolumns{5}
\tablewidth{0pt}
\tablecaption{Rotation Curve Fit Parameters}
\tablehead{\colhead{} & \colhead{} & \colhead{$\Phi_{\rm bulge}+\Phi_{\rm disk}+$} & \colhead{$\Phi_{\rm bulge}+\Phi_{\rm disk}+$} & \colhead{$\Phi_{\rm bulge}+\Phi_{\rm disk}+$}}
\startdata
Parameter & & $\Phi_{\rm caustic\; ring\; halo}$ & $\Phi_{\rm log \; halo}$ & $\Phi_{\rm NFW \; halo}$\\
\\[-0.4cm]
\\[-0.4cm]
disk mass & $M_{\rm disk}\;(M_{\odot})$ & $9.89\e{10}$ & $9.11\e{10}$ & $7.33\e{10}$\\
disk scale length & $a$ (kpc) & 5.0 & 5.0 & 5.0\\
disk scale height & $b$ (kpc) & 0.26 & 0.26 & 0.26\\
bulge mass & $M_{\rm bulge}\;(M_{\odot})$ & $1.89\e{10}$ & $1.89\e{10}$ & $1.89\e{10}$\\
bulge scale radius & $c$ (kpc) & 0.33 & 0.33 & 0.33\\
\\[-0.4cm]
 & $\chi^2_{v}$ & 1.04 & 1.01 & 1.19\\
\enddata
\label{rcfit}
\end{deluxetable}

In Figure \ref{rc}, the sum of a caustic ring halo (lower solid line), Hernquist bulge (dotted), Miyamoto-Nagai disk (dashed), HI gas layer (dashed-dotted line), and $\rm H_2$ gas layer (dashed-dotted line) produce a rotation curve (upper solid line) that matches observations. The blue data points are from \citet{2012PASJ...64...75S}. The data from \citet{2012PASJ...64...75S} is a compilation of previous rotation curve data from the literature with the velocities scaled to assume a rotation speed of 200 $\rm km\: s^{-1}$ at 8.0 kpc. The dashed-dotted line is a polynomial fit to the gas model (open circles for $\rm H_2$ and crosses for HI) from \citet{2000MNRAS.311..361O} as discussed above. The caustic ring halo rotation curve maintains kinks in the curve at the position of the caustics, similar to the acceleration plot (Figure \ref{accel}). Although the data and the model are similar, the observations are not detailed enough particularly for $r>10$ kpc to correlate bumps in the rotation curve with rises in the data points, as was suggested as evidence for the existence of caustic rings by \citet{2003PhLB..567....1S}.

\begin{figure}
\epsscale{0.7}
\plotone{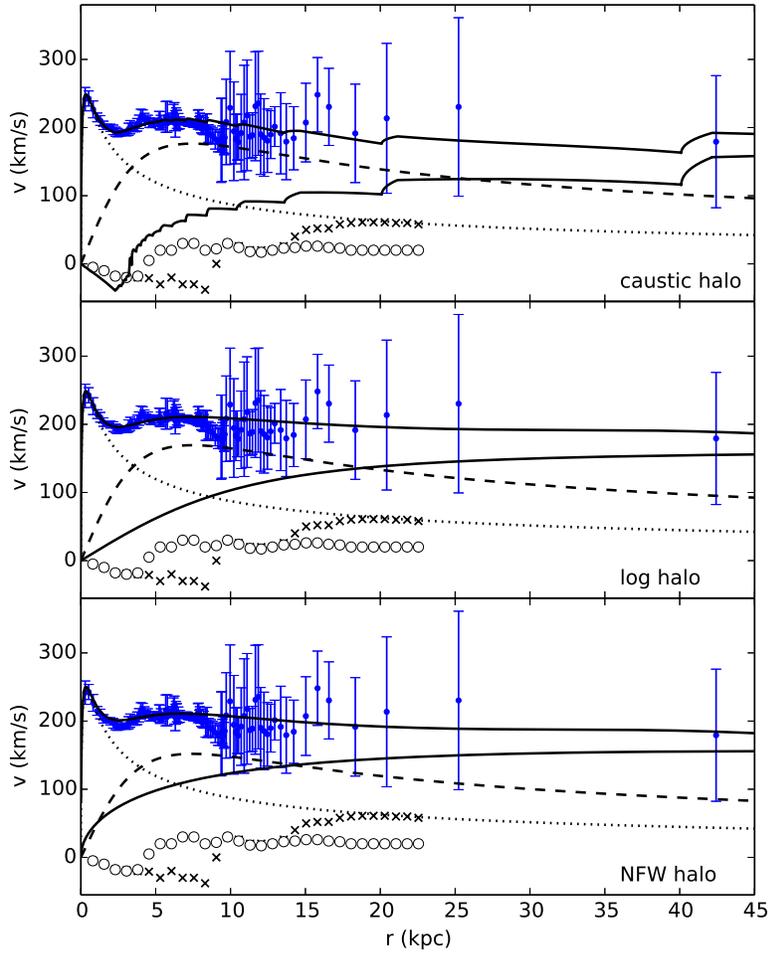}
\caption{Rotation curve of the Milky Way. Reproduction of Figure \ref{rc} with the logarithmic (middle) and NFW (bottom) halo shown for comparison. All three models are consistent with the measured rotation curve, but with different disk masses.}
\label{rc2}
\end{figure}

Figure \ref{rc2} shows rotation curves generated with the same disk, bulge, and gas components as Figure \ref{rc}, except with different halo models. The rotation curve fit for $0<r<2$ depends primarily on the contribution of the bulge model. The bulge mass and scale radius determines the fit to the region near the central peak at $r\sim0.3$. The fit for $2<r<15$ depends mostly on the contributions of the disk and halo. The caustic ring model has a negligible amount of mass within 2 kpc of the Galactic center and also does not contribute very much to the rotation curve for the inner Galaxy. This is shown in Figure \ref{rc2} (top panel) where the disk curve is much higher than the halo curve for $0<r<15$. The same is true for the case of a log halo model (middle panel). The NFW halo conversely predicts a much larger amount of mass for the inner Galaxy. In Figure \ref{rc2}, the NFW halo curve is almost level with the disk model curve. This exposes a degeneracy in the choice of the disk or halo model to fit the rotation curve data. Here the choice for a caustic ring halo or log halo are consistent with the maximum disk hypothesis \citep[discussed in][]{2008gady.book.....B}. This is the idea that the disk should contribute most (50-90\% of the total radial force) to the rotation curve of the inner Galaxy, and assumes that the mass-to-light ratio of the disk doesn't depend on the radius.   

The fit values for the disk and bulge parameters are shown in Table \ref{rcfit}. The main difference in the fits for the three panels in Figure \ref{rc2} is the value of the disk mass $(M_{\rm disk})$: 9.89\e{10}, 9.11\e{10}, and 7.33\e{10} $M_{\odot}$. In Table \ref{rcfit}, $a$, $b$, $M_{\rm bulge}$, and $c$ are identical for each column. Estimates for the Milky Way disk and bulge masses vary depending on which particular model and technique is used to fit the observational data. \citet{1998MNRAS.294..429D} used a Galaxy model with a spheroidal bulge, exponential disk, and spheroidal halo to fit values of $M_{\rm disk}\approx5.13\e{10}\; M_{\odot}$ and $M_{\rm bulge}\approx0.518\e{10}\; M_{\odot}$ for the disk and bulge. \citet{2005ApJ...631..838W} fit values of $M_{\rm disk}\approx4.58\e{10}\; M_{\odot}$ and $M_{\rm bulge}\approx1.1\e{10} \; M_{\odot}$ for a Galaxy model with a central black hole, Hernquist bulge, exponential disk, and NFW halo. \citet{2011MNRAS.414.2446M} fit a value of $M_{\rm disk}\approx5.76\e{10}\; M_{\odot}$ assuming a bulge mass of $M_{\rm bulge}=0.897\e{10}\; M_{\odot}$ for a Galaxy model with a \citet{2002MNRAS.330..591B} bulge, exponential disk, and NFW halo. \citet{2014ApJ...794...59K} fit values of $M_{\rm disk}\approx9.5^{+2.4}_{-3.0}\e{10}\; M_{\odot}$ and $M_{\rm bulge}\approx0.91^{+0.31}_{-0.38}\e{10}\; M_{\odot}$ for a Galaxy model with a spheroidal bulge, Miyamoto-Nagai disk, stellar halo, and NFW dark halo. The disk masses we found here are larger than what was found for an exponential disk but within the errors of what was found for a Miyamoto-Nagai disk. Our bulge masses are sightly larger than what was found in the sources listed here.

\section{The Sagittarius Dwarf Tidal Stream}

In this context, tidal streams are collections of stars that have been gravitationally stripped from dwarf galaxies that in turn are orbiting much larger galaxies. As a dwarf galaxy is tidally disrupted, two streams or ``tails'' of stars are formed. The leading tail moves ahead of the orbit of the bound dwarf galaxy and the trailing tail follows behind it. The Sagittarius (Sgr) dwarf galaxy \citep{1994Natur.370..194I} tidal stream is the largest and best studied tidal stream in the Milky Way galaxy \citep{2003ApJ...599.1082M, 2004ApJ...615..738M, 2005AJ....129..189V, 2005ApJ...619..800J, 2005ApJ...619..807L, 2010ApJ...714..229L}. 

The Sgr stream has been used to measure the Milky Way potential and thus constrain the shape of the dark matter halo. The data for Sgr tidal debris has however been difficult to fit to halo models. There have been claims for spherical, prolate, and oblate ellipsoid models. Since the Sgr debris is approximately confined to a plane, \citet{2003ApJ...599.1082M} suggested that the potential should be spherical. \citet{2005AJ....129..189V} compared radial velocities and distances from the leading tidal tails to numerical simulations for different halo models. They found that prolate and spherical models fit better but still fail to reproduce the entire angular span of the stream. The tilt of the tidal tails with respect to the Sgr orbital plane, however, suggests that the potential is oblate \citep{2005ApJ...619..800J}. \citet{2005ApJ...619..807L} conversely found the prolate models best fit the leading tail velocities. More recently, \citet{2010ApJ...714..229L} found a reasonable fit to both tidal tails if they assume a triaxial halo, with both the major and minor axes in the plane of the Milky Way. \citet{2013ApJ...773L...4V} expanded upon this idea to fit a halo potential that varies as a function of Galactocentric radius. These authors obtain an oblate fit within 10 kpc of the Galactic center and a fit to the triaxial model for outer radii. Due to the controversy in modeling the tidal tails of the Sgr stream, it is worth asking whether substructures such as caustic rings could affect the shape and velocity structure of this tidal debris stream.

As a test of the viability of the caustic ring model, we determine whether Sgr stream observations are consistent with orbits and N-body simulations in a caustic ring halo potential. In Section 5.1, we discuss the orbit fitting technique used, and the results of comparing the model to the observational data. After the orbits are fit, N-body simulations are run with those orbits. We will compare the N-body models to the data in Section 5.2, using three cases for the fixed analytical potential of the Milky Way:
\begin{enumerate}
\item
$\rm \Phi_{Galaxy}= \Phi_{Hernquist \; bulge}+\Phi_{Miyamoto-Nagai \; disk}+\Phi_{caustic \; halo}$
\item
$\rm \Phi_{Galaxy}= \Phi_{Hernquist \; bulge}+\Phi_{Miyamoto-Nagai \; disk}+\Phi_{NFW \; halo}$
\item
$\rm \Phi_{Galaxy}= \Phi_{Hernquist \; bulge}+\Phi_{Miyamoto-Nagai \; disk}+\Phi_{triaxial \; halo}$
\end{enumerate}
The galaxy models used here are those that have been fit to the Milky Way rotation curve in Section 3. The disk, bulge, and halo parameters are from Table \ref{halomodel} and Table \ref{rcfit}. For the case of the triaxial halo potential, we use the disk and bulge parameters from the log halo fit in Table \ref{halomodel} and Table \ref{rcfit}.

\subsection{Orbit Fitting}
\begin{figure}
\plotone{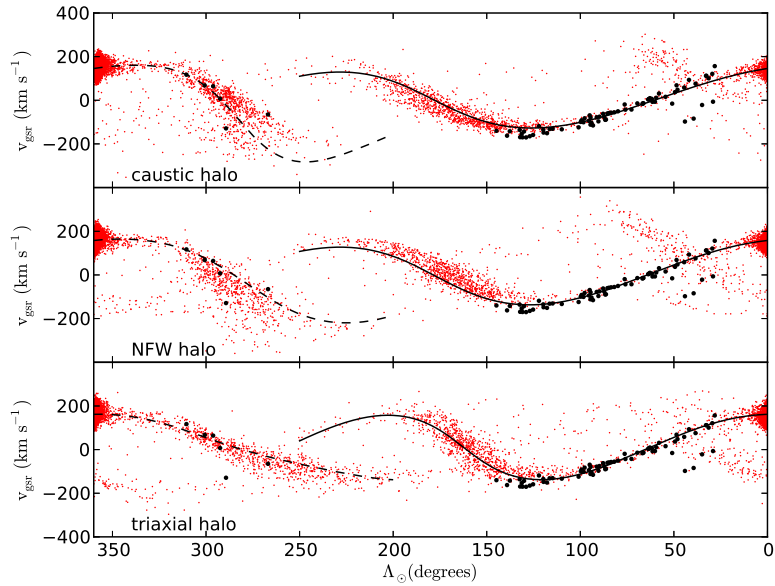}
\caption{Galactic standard of rest velocities vs. orbital longitude (Sgr heliocentric spherical coordinate system) for the Sgr dwarf galaxy tidal disruption. The forward (dotted line) and backwards (solid line) best-fit orbit to the 2MASS M giant data from \citet{2003ApJ...599.1082M} (black dots) is shown overplotted with their corresponding N-body simulations (red dots). The orbits and N-body fits for the caustic (top), NFW (middle), and triaxial (bottom) halo are shown for comparison.}
\label{vL}
\end{figure} 

\begin{figure}
\plotone{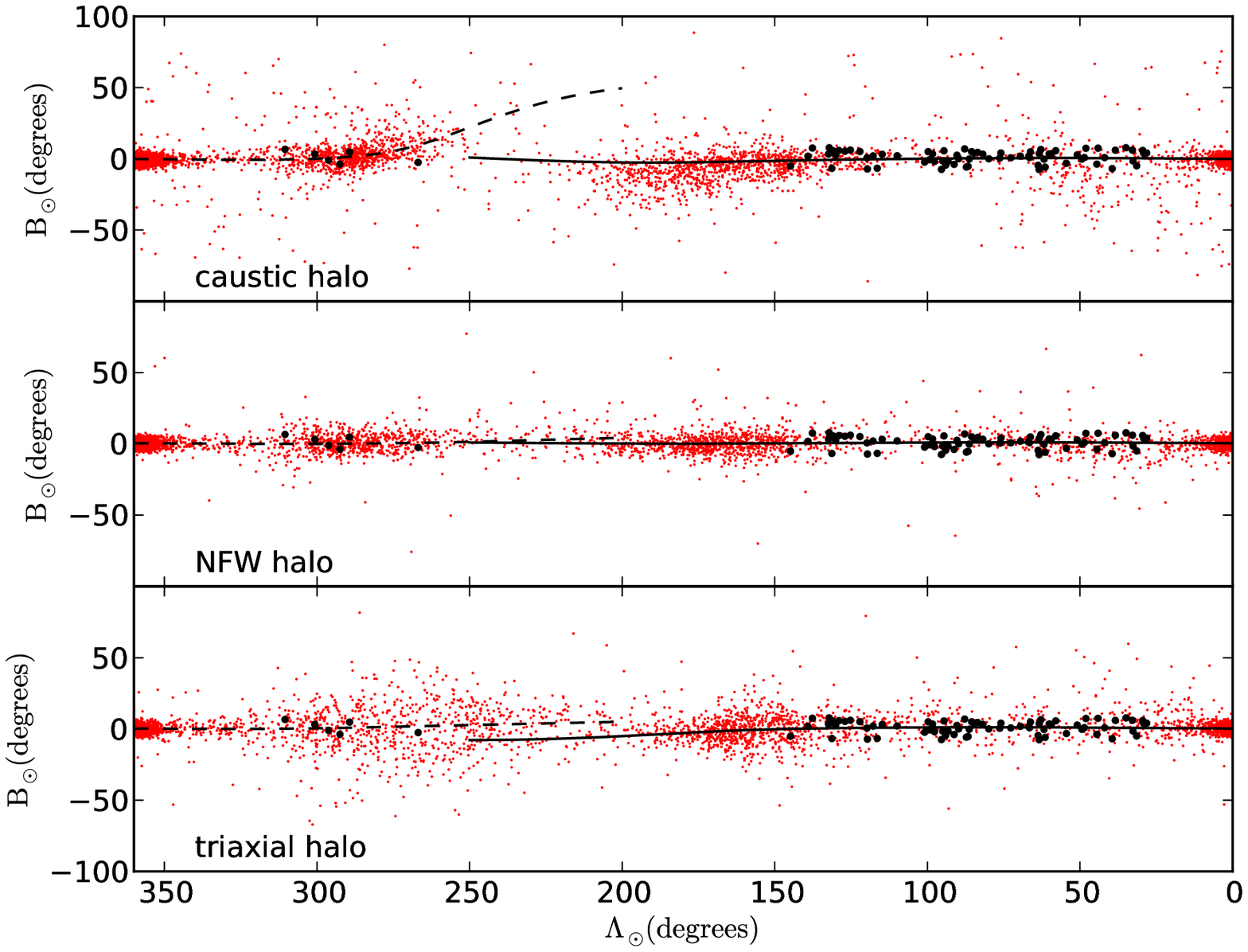}
\caption{Orbital latitude vs. orbital longitude (Sgr heliocentric spherical coordinate system) for the Sgr dwarf galaxy tidal disruption. The forward (dotted line) and backwards (solid line) best-fit orbit to the 2MASS M giant data from \citet{2003ApJ...599.1082M} (black dots) is shown overplotted with their corresponding N-body simulations (red dots). The orbits and N-body fits for the caustic (top), NFW (middle), and triaxial (bottom) halo are shown for comparison.}
\label{BL}
\end{figure} 

An orbit (line through phase space) of a test particle through a particular external potential is specified by initial conditions for position and velocity. We generate orbits using the {\it mkorbit} and {\it orbint} tools from the NEMO Stellar Dynamics Toolbox \citep{1995ASPC...77..398T}. The mkorbit tool generates an orbit file given a specified analytical external potential and starting values for position and velocity. The {\it orbint} tool integrates the orbit through a specified number of timesteps. We use the orblist tool to list the position and velocity values calculated at each timestep. 

We can constrain the potential of the Milky Way by matching model orbits with observational data from stars in tidal streams. We used a gradient descent method to fit Sgr stream data to orbits using methods described in section 6 of \citet{2009ApJ...697..207W}. The NEMO code generally uses Galactocentric Cartesian coordinates for input of positions and velocities $(x,\; y, \; z, \; v_x, \; v_y, \; v_z)$. For convenience, we will use the Sgr heliocentric coordinate system defined in \citet{2003ApJ...599.1082M} to designate positions $(d_{\rm sun}, \; \Lambda_{\odot}, \; B_{\odot})$. In this section and the next we adopt a distance of $R_{\odot}=8.0$ kpc from the Sun to Galactic center. Now the initial conditions for an orbit are specified by: $(d_{\rm sun}, \; \Lambda_{\odot}, \; B_{\odot}, \; v_x, \; v_y, \; v_z)$. 

We use the reduced $\chi^2$ as measure of fitness of a model orbit to stream data points:
\begin{equation}
\chi^2_{v_{\rm gsr}}= \sum\limits_{i} \left(\frac{v_{{\rm gsr\; model},i}-v_{{\rm gsr\; data},i}}{\sigma_{v_{\rm gsr},i}}\right)^2,
\end{equation}
\begin{equation}
\chi^2_{B}= \sum\limits_{i} \left(\frac{{B_{{\rm\; model},i}}-{B_{{\rm data},i}}}{\sigma_{{B},i}}\right)^2,\; \rm and
\end{equation}
\begin{equation}
\chi^2=\frac{1}{N-n-1}(\chi^2_{v_{\rm gsr}}+\chi^2_{B}).
\end{equation}
where $v_{\rm gsr}$ is the Galactic standard of rest velocity, $\sigma$ is the error in the measurement of the data point, $N$ is the number of data points, and $n$ is the number of parameters being fit. Since the orbit is a line through phase space, the best fit is one where the orbit line goes mostly closely through the set of stream data points. The fit is therefore determined by the starting positions and velocities and the choice of the external Galactic potential model. Here we can therefore compare orbit fits in different external potentials. 

Since the current coordinates of the Sgr dwarf galaxy are well known, we fix the starting points for the distance and orbital longitude and fit the starting points for the orbital latitude and velocity $(d_{\rm sun}, \; \Lambda_{\odot}, \; B_{\odot}, \; v_x, \; v_y, \; v_z)=(28.0 \; {\rm kpc}, \; 0.126^{\circ}, \; B_{\odot}, \; v_x, \; v_y, \; v_z)$. We are fitting 4 parameters. The orbit was fit to 2MASS M giant angular position/velocity data from Table 3 of \citet{2004ApJ...615..738M}. Only M giants with $\rm d_{\rm sun}>13$ kpc and $-7.8<B_{\odot}<7.8$ were selected for our fits. These points were split into $10^{\circ}$ bins in latidude $\Lambda_{\odot}$, and points with velocities greater than $3\, \sigma$ of the bin average were removed from the data set. The error of each velocity point was taken to be $\sigma_{v_{\rm gsr},i}=6 \; \rm km \: s^{-1}$ as stated in \citet{2004ApJ...615..738M}. The error in latitude, $\sigma_{B,i}=3$ is the standard deviation of $B_{\odot}$ in our data set. The number of data points is $N=80$ and the number of parameters is $n=4$. We completed orbit fits for the three cases listed above for the external Galactic potential, which highlights the effects of varying the choice of the halo model. 

The gradient descent method takes an initial set of parameters $(d_{\rm sun}, \; \Lambda_{\odot}, \; B_{\odot}, \; v_{x1}, \; v_{y1}, \; v_{z1})$, generates an orbit, compares the velocities from this orbit to Sgr M giant data, and calculates a $\chi^2$ value, $\chi^2_{\rm old}$. Next the velocities are adjusted to new values $(d_{\rm sun}, \; \Lambda_{\odot}, \; B_{\odot}, \; v_{x2}, \; v_{y2}, \; v_{z2})$ by moving in the direction of the $\chi^2$ gradient, a new orbit is generated and $\chi^2_{\rm new}$ is calculated. If $\chi^2_{\rm new}$ is less than $\chi^2_{\rm old}$, the new set of parameters are kept. If $\chi^2_{\rm new}$ is greater than $\chi^2_{\rm old}$, the old set of parameters are kept. The process is repeated until $\chi^2_{\rm new}=\chi^2_{\rm old}$, signaling that a minimum has been reached. The best fit parameters, $(d_{\rm sun}, \; \Lambda_{\odot}, \; B_{\odot}, \; v_x, \; v_y, \; v_z)=
(28.0 \; {\rm kpc}, \; 0.126^{\circ}, \; B_{\odot, \; {\rm best fit}}, \; v_{x, \; {\rm best fit}}, \; v_{y, \; {\rm best fit}}, \;$ $v_{z, \; {\rm best fit}})$
correspond with a minimum $\chi^2$ value. The errors in the fit parameters are estimated using the method described in \citet{2008ApJ...683..750C}. We assume the shape of the $\chi^2$ surface near the minimum is Gaussian. The curvature of a surface is found using the Hessian or curvature matrix $\mathbf{H}$, a matrix containing all possible second derivatives of a function. We calculate the derivatives for the Hessian matrix numerically using Equation 23 from \citet{2008ApJ...683..750C}. The inverse of the Hessian matrix gives the error with respect to each best fit parameter:
\begin{equation}
\mathbf{V}=\frac{1}{N}\mathbf{H^{-1}}=\sigma^2.
\end{equation}
The diagonal elements of $\mathbf{V}$ are the variances and the off diagonal elements are the covariances. The variance is normalized by the number of data points N. $\sigma_i=\sqrt{V_{ii}/N}$ gives the error estimate with respect to each best fit parameter.

The starting velocities for the gradient descent were random values chosen from a range: $(v_{x}, \; v_{y}, \; v_{z})=$ (150 to 250, -40 to 20, 150 to 250). These values were chosen to ensure the gradient descent algorithm begins close the orbit of Sgr that is known from previous publications. The gradient descent however will not necessarily end with values in these ranges. A disadvantage to using the gradient descent method is that if the starting point is very far from the global minimum, the algorithm may become trapped in a local minimum. We start with these values to ensure the global minimum is found. Our best fits are chosen as the run with the lowest $\chi^2$ value after 10 separate runs. The results for the fits in the 3 cases listed above are: $(B_{\odot,\;\rm best fit}, \; v_{x, \; {\rm best fit}}, \; v_{y, \; {\rm best fit}}, \; v_{z, \; {\rm best fit}})=$

\begin{enumerate}
\item
$\rm (0.028^\circ \pm 0.7,\; 210.2 \pm1.1 \; km \; s^{-1}, \; -30.8 \pm1.1 \; km \; s^{-1}, \; 212.7 \pm0.7 \; km \; s^{-1})$, $\; \chi^2=52.2 \;$  (caustic)
\item
$\rm (0.622^\circ \pm 0.7,\; 222.9 \pm1.4 \; km \; s^{-1}, \; -32.8 \pm1.4 \; km \; s^{-1}, \; 208.7 \pm0.8 \; km \; s^{-1})$, $\; \chi^2=45.8 \;$ (NFW)
\item
$\rm (0.441^\circ \pm 0.7,\; 224.6 \pm1.4 \; km \; s^{-1}, \; -29.4 \pm1.4 \; km \; s^{-1}, \; 198.7 \pm0.9 \; km \; s^{-1})$, $\; \chi^2=53.7 \;$ (triaxial)
\end{enumerate}

Figures \ref{vL} and \ref{BL} show the best fit orbits to the M giant data (black points) for cases 1-3. The trailing data lies on the backwards orbit (solid line) and the leading tail data lies on the forwards orbit (dotted line). We have plotted the Galactic standard of rest velocity $(v_{\rm gsr}=x \cdot v_x+y \cdot v_y+z \cdot v_z)$ vs. orbital longitude $\Lambda_{\odot}$ in Figure \ref{vL}. The angle along the tidal stream, $\Lambda_{\odot}$, is zero at the position of the Sgr dwarf galaxy and defined to increase in the direction of trailing debris \citep{2003ApJ...599.1082M}. The $\chi^2$ values for each fit are on the order 50 because of the outliers in the data from $\Lambda_{\odot}=0^\circ-50^\circ$ and the spread in range of the leading tail velocity points. The two data points on the edge of the leading tail are about 50-100 $\rm km\: s^{-1}$ away from the orbit fit lines, yet from visual inspection of the plot the fits in the three panels are reasonably close to the data points. The trailing tail orbits in all three potentials are similar. In the leading tail, the orbits in NFW and triaxial halos favor upper data points. Yet the caustic ring halo orbit moves through the center of the points. Since the $\chi^2$ values for all three fits are similar, perhaps more data is needed in the trailing arm to determine which fit is favored. We have also plotted the orbital latitude, $B_{\odot}$, vs. orbital longitude, $\Lambda_{\odot}$, in Figure \ref{BL}. $B_{\odot}\approx0$ is in the plane of the Sgr dwarf galaxy orbit. The orbit fits in each panel are all well confined to the Sgr plane. The N-body simulations (red points) in Figures \ref{vL} and \ref{BL} are discussed in the next section. 

We estimate the radial period of the best fit orbits by calculating the time difference in going from the first backwards orbit apocenter to the first forwards orbit apocenter. We found the radial periods to be 1.09 Gyr (caustic), 1.18 Gyr (NFW), and 0.92 Gyr (triaxial). These values are similar to those found in \citet{2005ApJ...619..807L}: 0.85 Gyr, 0.88 Gyr, and 0.87 Gyr for oblate, spherical, prolate potentials, respectively.

\subsection{N-body simulations of the Sagittarius dwarf tidal stream}

In this section, we will compare the results of N-body simulations of the tidal disruption of the Sgr dwarf galaxy. We use the {\it gyrfalcON} tool \citep{2000ApJ...536L..39D, 2001MNRAS.324..273D, 2002JCoPh.179...27D} in the NEMO Stellar Dynamics Toolbox \citep{1995ASPC...77..398T} to run the N-body simulations. Dwarf galaxies were modeled by a Plummer sphere \citep{1911MNRAS..71..460P}:
\begin{equation}
\Phi_{\rm P}=-\frac{GM_{\rm P}}{\sqrt{r^2+r_o^2}}
\label{plummer}
\end{equation}
with initial mass $M_{\rm P}=M_{\rm Sgr,0}=1.8\e{9}\;M_{\odot}$ and scale length $r_o=1.0$ kpc with the number of bodies $\rm N=10^4$. We assume the initial mass is the sum of the light and dark matter components of the dwarf galaxy. The caustic ring model does not specifically address the dark matter distribution in dwarf galaxies, but it is reasonable to assume that, like the Milky Way, they would not include a central cusp as in an NFW profile. We use $M_{\rm Sgr,0} \sim 1.8\e{9}\;M_{\odot}$, which is slightly larger than the value of $M_{\rm Sgr,0} \sim 6.4\e{8}\;M_{\odot}$ used by \citet{2010ApJ...714..229L} but smaller than $10^{10}-10^{11}\;M_{\odot}$ found by \citet{2010ApJ...712..516N} and used by \citet{2011Natur.477..301P}. The exact initial mass of the Sgr dwarf galaxy is not important to the results of our simulation.

\begin{figure}
\plotone{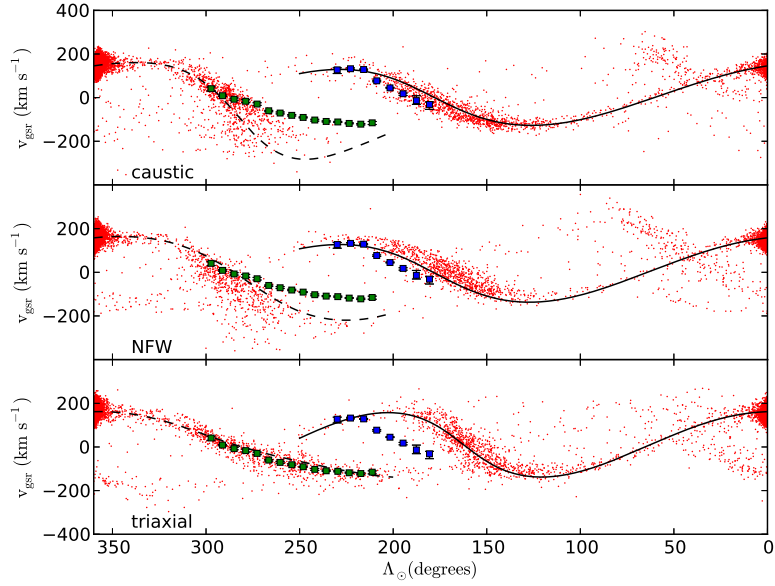}
\caption{Galactic standard of rest velocities vs. orbital longitude for N-body simulations of the Sgr dwarf galaxy tidal disruption. Figure \ref{vL} is reproduced with recent data for the trailing (blue squares) and leading (green squares) tidal tail from \citet{2014MNRAS.437..116B}. The error bars are shown but most of these are smaller than the size of the squares.}
\label{vL2}
\end{figure}

\begin{figure}
\epsscale{0.68}
\plotone{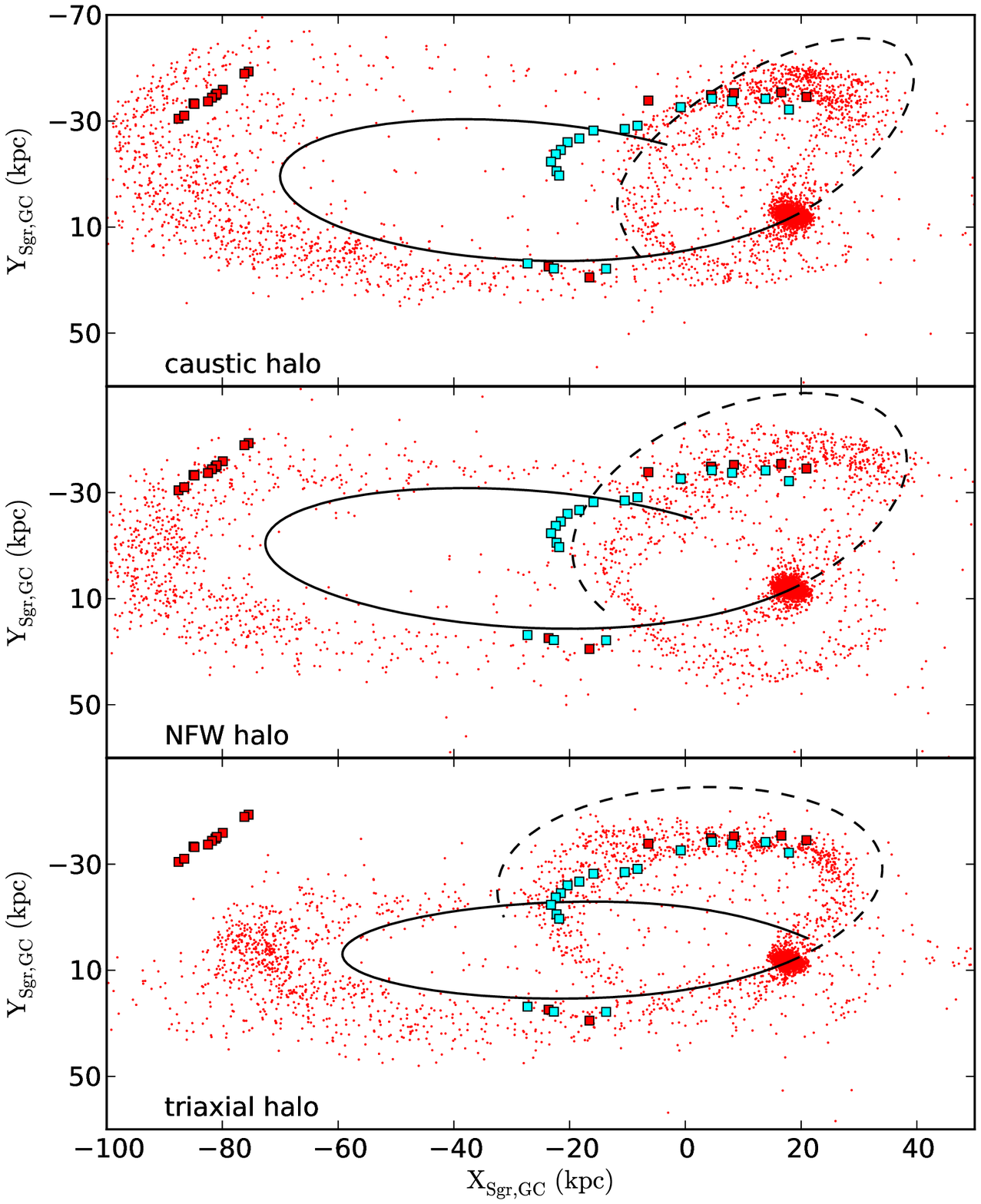}
\caption{Ysgr,gc vs. Xsgr,gc (Sgr Galactocentric spherical coordinate system) for N-body simulations of the Sgr dwarf galaxy tidal disruption. The blue dots show the galaxy evolved in a caustic (top), NFW (middle), and triaxial (bottom) halo potential. The forward (dotted line) and backwards (solid line) orbit is also plotted. Recent data from \citet{2013AJ....145..163N} (cyan squares) and \citet{2003ApJ...596L.191N} (red squares) is also shown for comparison. Note both the caustic ring halo and NFW halo tidal debris extends as far as the 90 kpc data, but the triaxial halo fits the leading tidal tail.}
\label{xygc}
\end{figure}

The Plummer model is placed at a point along the backwards orbit (in the past) and evolved to the position of the dwarf galaxy at the present day. The model is evolved in an external Galactic potential corresponding to the three cases listed in Section 5.1. The fixed parameters for the potential are also the same as listed above. The simulations in Figures \ref{vL}-\ref{xygc} were run for a time equal to three orbital periods in the respective potentials: 3.28 Gyr (caustic), 3.54 Gyr (NFW), and 2.77 Gyr (triaxial). 

In Figures \ref{vL}-\ref{xygc}, the Plummer model (red points) is evolved along the best fit orbits determined in the previous section. The panels show the results in the three halo potentials: caustic (top), NFW (middle), and triaxial (bottom) as a function of the angle along the tidal stream, $\Lambda_{\odot}$. Figure \ref{vL} shows $v_{\rm gsr}$ vs. $\Lambda_{\odot}$. The caustic ring halo plot has much lower velocities at the end of the leading (dashed line) tidal tail, and one can even imagine that the two M giant stars are each from a different apparent branch of the N-body simulation. The N-body models in all three halos match the trailing tail data well. The caustic and NFW debris matches the leading tail data well, whereas the triaxial debris matches five of the data points and misses the sixth one. More data points however would be needed to determine which fit is preferred in the leading arm. Figure \ref{BL} shows $B_{\odot}$ vs. $\Lambda_{\odot}$. The N-body models in all 3 potentials are well confined to the Sgr orbital plane.

In Figures \ref{vL2} and \ref{xygc} we compare the N-body simulations to a set of more recent Sgr data that extends farther along the leading and trailing tidal tails, in the range $150<\Lambda_{\odot}<250$. Figure \ref{vL2} shows the simulations overplotted with recent velocity data for giant stars detected photometrically from SDSS by \citet{2014MNRAS.437..116B}. Here the caustic and NFW debris match the trailing data (blue points) but the triaxial debris does not. Conversely, the triaxial debris matches the leading tail data (green points), but the NFW and caustic debris doesn't for the points from $\Lambda_{\odot}=200^\circ-250^\circ$. Figure \ref{xygc} is plotted using the Sgr Galactocentric Cartesian coordinate system defined in \citet{2003ApJ...599.1082M}. The center of the Sgr Galactocentric system is the point in the Sgr dwarf galaxy plane that is closest to the Galactic center. Because the orbit of the Sgr dwarf galaxy is nearly polar, $\rm (X_{sgr,gc}, Y_{sgr,gc}, Z_{sgr,gc})$ are approximately in the directions (X, -Z, Y), where X points from the Sun to the Galactic center, Y is in the direction of the Sun's motion, and Z in the direction of the North Galactic Pole. The simulation in Figure \ref{xygc} is overplotted with recent Sgr distance observations from \citet{2003ApJ...596L.191N} and \citet{2013AJ....145..163N}. \citet{2003ApJ...596L.191N} measured the distances of BHB stars detected photometrically from SDSS. \citet{2013AJ....145..163N} measured Sgr stream center distances by fitting the spatial density of F turnoff stars in SDSS using the statistical photometric parallax technique. The caustic and NFW N-body models extend as far as 90 kpc from the Galactic center, much further than the triaxial N-body model. This is plausible because the caustic ring halo potential well is shallow (see Figure \ref{accel}), and the bodies can maintain enough kinetic energy to fly out to further distances. The caustic and NFW debris however doesn't match the data for the leading arm positions. The triaxial debris is the best fit to the leading arm. We find here that none of the models are able to match both the leading and trailing tail data simultaneously.

The orbital precession of the stream is found by calculating the change in the angle $\Lambda_{\odot}$ between the leading and trailing apocenters. Since the precession depends on gravitational potential, it is worth observing how it varies in the three potentials we compare in this section. We fit 2nd order polynomials to the N-body model debris and then found the angle between the apocenters. The precesssion values are $\delta\Lambda_\odot=103.6^\circ$ (caustic), $\delta\Lambda_\odot=124.2^\circ$ (NFW), and $\delta\Lambda_\odot=132.9^\circ$ (triaxial). In the Galactocentric coordinate system \citep{2003ApJ...599.1082M} the values are $\delta\Lambda_{GC}=100.8^\circ$ (caustic), $\delta\Lambda_{GC}=114.2^\circ$ (NFW), and $\delta\Lambda_{GC}=123.1^\circ$ (triaxial). \citet{2014MNRAS.437..116B} found values of $\delta\Lambda_\odot=99.3\pm3.5^\circ$ and $\delta\Lambda_{GC}=93.2^\circ\pm3.5^\circ$ from detections of BHB stars from SDSS in the leading and trailing tails. The caustic halo values are closest to those found by \citet{2014MNRAS.437..116B}. \citet{2014MNRAS.437..116B} show that for spherical potentials the precession angle increases as the rotation curve gets steeper. We can assume the caustic ring halo potential is approximately spherical since the curve along the radial and $z$ directions have the same scale and general shape (see Figure \ref{accel}). By inspection of Figure \ref{rc2}, it makes sense that the precession angle in the NFW halo is larger than that in the caustic ring halo since the NFW rotation curve is steeper. 

In summary, we show that N-body simulations of the Sgr stream in a caustic ring halo match most of the recent data for stream velocities and distances except on the edge of the leading tail at $\Lambda_{\odot}=200^\circ-250^\circ$. Neither of the caustic, NFW, or triaxial models are able to match velocity/distance data from the leading and trailing tails simultaneously. Lastly, the caustic halo N-body model can match Sgr distance observations in the trailing tidal tail, including debris out to 90 kpc.

\section{The Effect of a Dark Matter Caustic Ring on a Satellite Passing Through It}

\begin{figure}
\plotone{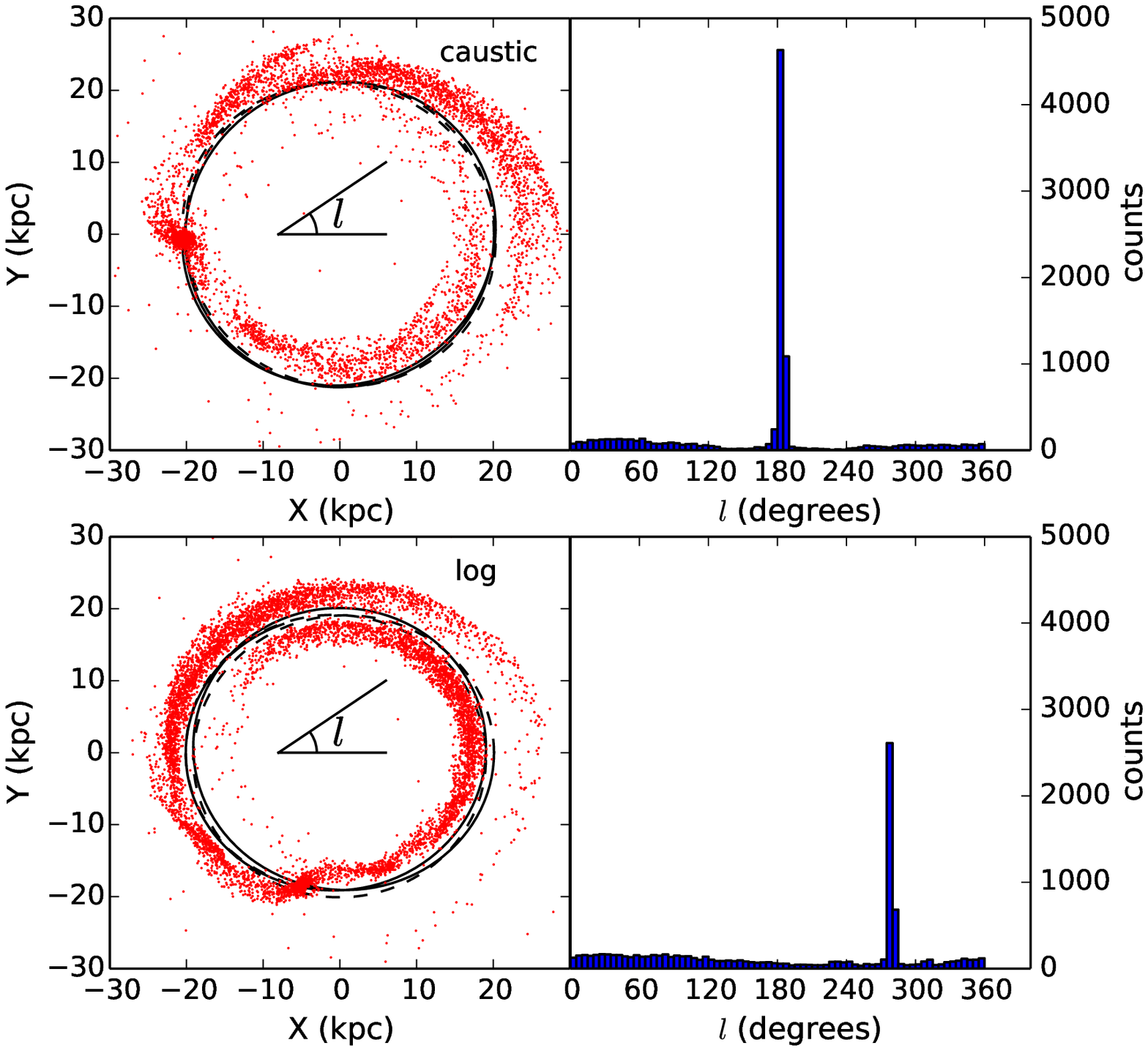}
\caption{$Y$ vs. $X$ (Galactocentric Cartesian coordinates) for N-body simulations of a dwarf galaxy that moves along an orbit in the $+Y$ direction when it is at the position of the $n=2$ caustic ring at $(X, Y)$=(20.1, 0) kpc in a log (bottom left panel) and caustic (top left panel) ring halo potential. The simulations were run for 2 Gyr. The forward (dotted line) and backward (solid line) orbit is also plotted. The right panels show histograms of the number of bodies as a function of Galactic longitude $l$.}
\label{r1}
\end{figure} 

\begin{figure}
\plotone{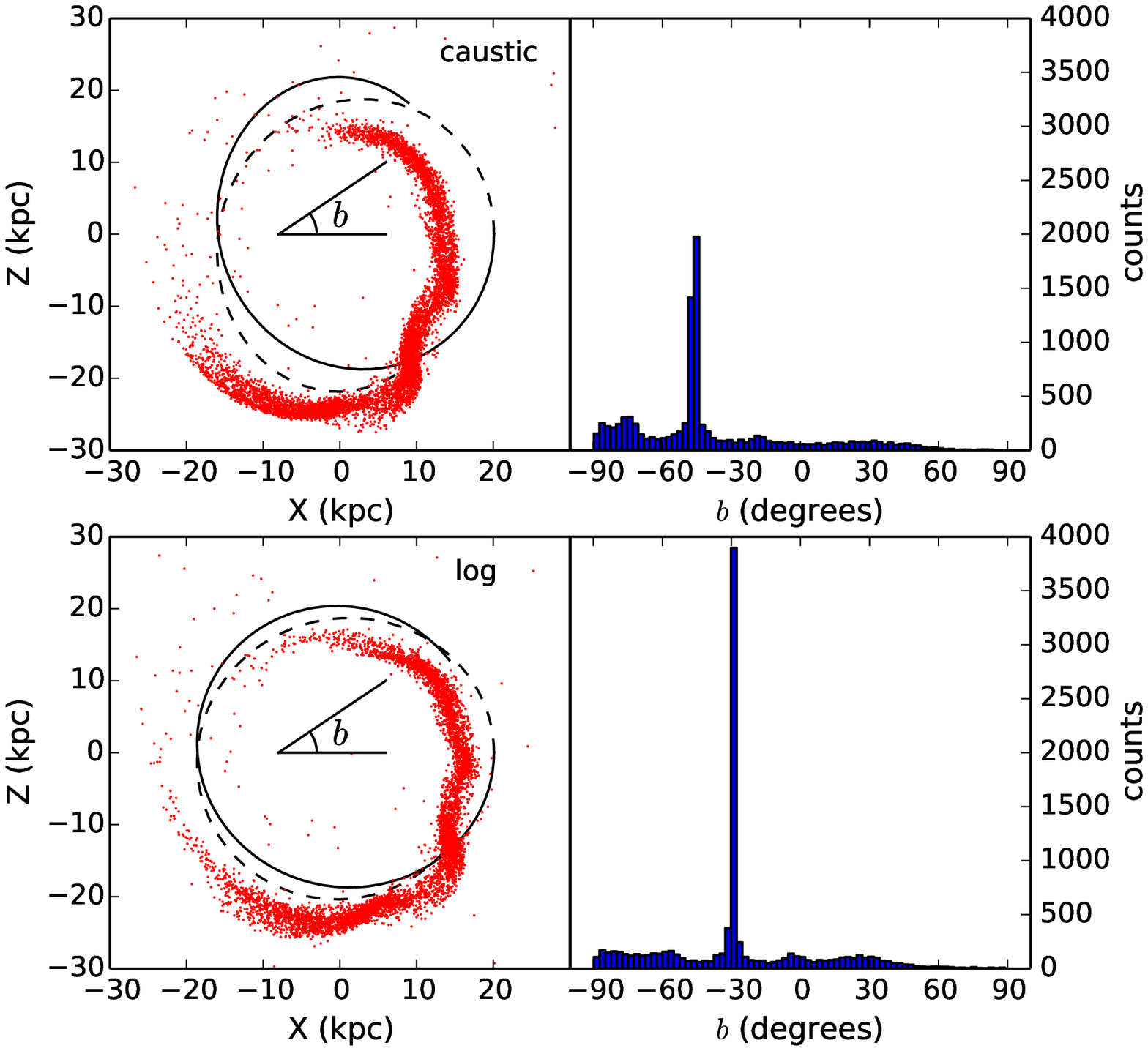}
\caption{$Z$ vs. $X$ (Galactocentric Cartesian coordinates) for N-body simulations of a dwarf galaxy that moves along an orbit in the $+Z$ direction when it is at the position of the $n=2$ caustic ring at $(X, Z)$=(20.1, 0) kpc in a log (bottom left panel) and caustic ring (top left panel) halo potential. The simulations were run for 1.1 Gyr. The forward (dotted line) and backward (solid line) orbit is also plotted. The right panels show histograms of the number of bodies as a function of Galactic latitude $b$.}
\label{r2}
\end{figure} 

\begin{figure}
\plotone{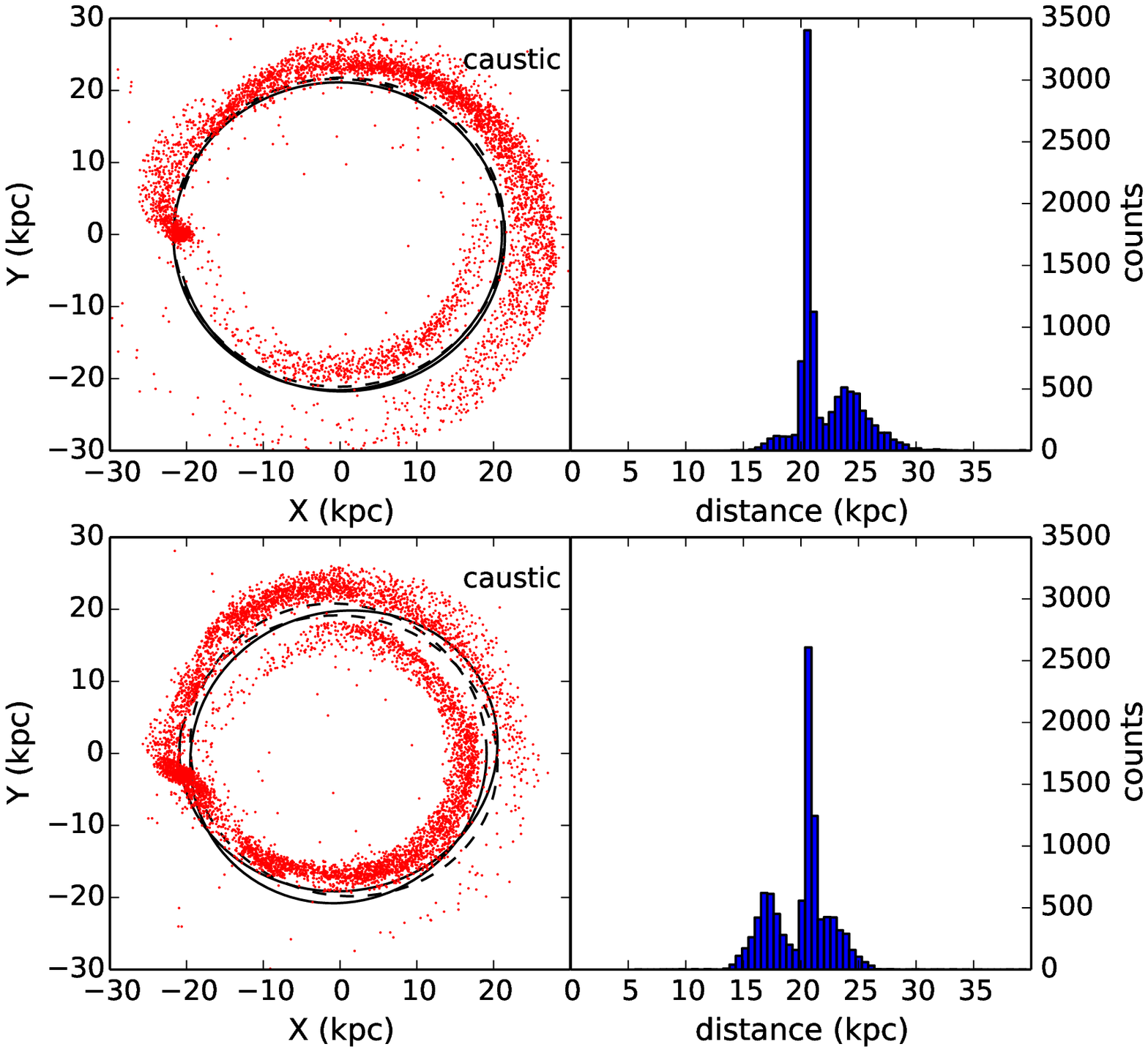}
\caption{$Y$ vs. $X$ (Galactocentric Cartesian coordinates) for N-body simulations of a dwarf galaxy that moves along an orbit in the $+Y$ direction when it is 1 kpc away from the $n=2$ caustic ring at $(X, Y)$=(21.1, 0) kpc (top left panel) or $(X, Y)$=(19.1, 0) kpc (bottom left panel) in a caustic ring halo potential. The simulations were run for 2 Gyr. The forward (dotted line) and backward (solid line) orbit is also plotted. The right panels show histograms of the number of bodies as a function of Galactocentric distance. Note that there is an excess of of debris at distances greater than 20 kpc in the top right histogram and an excess of of debris at distances less than 20 kpc in the bottom right histogram.}
\label{r3}
\end{figure}

We ran N-body simulations of dwarf galaxies represented by Plummer sphere models on orbits that pass close to caustic rings, to see whether the high density areas would have a strong effect on tidal disruption. We found that although there is a large discontinuity in the acceleration near a caustic, the caustic ring gravitational field does not qualitatively affect the results of the simulation. There are, however, differences in the shapes of the orbits and resulting tidal tails in a caustic ring halo potential compared to the spherical log halo potential. In Figures \ref{r1}-\ref{r3}, a Plummer model (Equation \ref{plummer}) dwarf galaxy with $M_P=10^8\;M_{\odot}$, $r_o=0.5$ kpc, and number of bodies $\rm N=10^4$, is evolved in two different Milky Way gravitational potentials:
\begin{enumerate}
\item
$\rm \Phi_{Galaxy}= \Phi_{Hernquist \; bulge}+\Phi_{Miyamoto-Nagai \; disk}+\Phi_{caustic\; halo}$
\item
$\rm \Phi_{Galaxy}= \Phi_{Hernquist \; bulge}+\Phi_{Miyamoto-Nagai \; disk}+\Phi_{log \; halo}$.
\end{enumerate}
The parameters for these models are listed in Table \ref{halomodel} and Table \ref{rcfit}.

In Figures \ref{r1} and \ref{r2}, the Plummer sphere is evolved along different orbital paths that move near the $n=2$ caustic ring at 20.1 kpc from the Galactic center (left boundary of tricusp in Figure \ref{g-ring}). The simulations start along the backwards orbit (solid line) and evolve to move along the caustic ring and then move along the forward orbit (dotted line). For Figure \ref{r1}, the orbit is designed to move in the $+Y$ direction through the side of the caustic ring in the plane of the Galaxy at $(X,\;Y,\;Z,\;v_{x},\;v_{y},\;v_{z})=(\rm 20.1\;kpc,\;0\;kpc,\;0\;kpc,\;0\;km \; s^{-1},$ $\rm 200.0\;km \; s^{-1},\;0\;km \; s^{-1})$. $X$, $Y$, and $Z$ are Galactocentric Cartesian coordinates with the position of the Sun taken as $R_{\odot}=8.0$ kpc or $(X,\;Y,\;Z)=(-8.0,\;0,\;0)$ kpc. The simulation time is 2 Gyr. The orbit in a caustic ring halo (top left panel) is approximately circular while the log halo orbit (bottom left panel) has some precession. The dwarf galaxies also travel to different positions after 2.0 Gyr. When we look at histograms of the number of bodies along Galactic longitude ($l$ is convenient since we are looking in the Galactic plane) we find that the dwarf less disrupted in a caustic halo. The central peak in a caustic halo has about 50\% of the total number of bodies and the log halo has about 30\% of the total. We also ran simulations where the Plummer sphere ended in the same position for the two potentials; those are not shown here. In this case we found the same behavior where the dwarf galaxy disrupts more slowly in a caustic potential. Here the dwarf galaxy may be on a stable orbit around the ring and the bodies are stripped from the galaxy core more slowly.  

In Figure \ref{r2}, the orbit is along a polar path and moves in the $+Z$ direction through the side of the caustic at $(X,\;Y,\;Z,\;v_{x},\;v_{y},\;v_{z})=$ $(\rm 20.1\;kpc,\;0\;kpc,\;0\;kpc,\;0\;km \; s^{-1},\;0\;km \; s^{-1},\;200.0\;km \; s^{-1})$, normal to the Galactic plane. The simulation time is 1.1 Gyr. The dwarf galaxy moves to almost the same position after 1.1 Gyr in both the caustic ring and log potentials. In this case the dwarf disrupts more slowly in log halo; the central peak in a caustic halo has about 20\% of the total number of bodies and the log halo has about 40\% of the total. The caustic ring may have the effect of stripping more bodies from the dwarf galaxy in its passage through the disk.

In Figure \ref{r3}, the orbit is designed to move in the $+Y$ direction of the Galactic plane just inside and outside of the caustic at $(X,\;Y,\;Z,\;v_{x},\;v_{y},\;v_{z})=(\rm 19.1\;kpc,\;0\;kpc,\;0\;kpc,\;0\;km \; s^{-1},\;200.0\;$ $\rm km \; s^{-1},\;0\;km \; s^{-1})$ and $(X,\;Y,\;Z,\;v_{x},\;v_{y},\;v_{z})=(\rm 21.1\;kpc,\;0\;kpc,\;0\;kpc,\;0\;km \; s^{-1},\;200.0\;km \; s^{-1},\;$ $\rm 0\;km \; s^{-1})$. These simulations were run for 2 Gyr. Here an interesting result is found. If the orbit is started just inside the $n=2$ caustic ring, most of the stars are disrupted interior to the rings with distances less that 20 kpc. If we start just outside the ring the tidal debris is concentrated greater then 20 kpc just outside the ring.

In summary, we find that the caustic rings do have subtle influences on the formation of tidal tails, but the differences are not dramatic. The study in this section is motivated by the proposal that the Monoceros Ring, a ring of stars about 20 kpc from the Galactic center, is associated with a dark matter caustic ring at 20 kpc (\citealt{2007PhRvD..76b3505N}, citing \citealt{2002ApJ...569..245N}). Certainly we designed the orbits in Figures \ref{r1}-\ref{r3} to move along a caustic ring to study its effects, but realistically dwarf galaxies would not necessarily fall in to the Milky Way with such prescribed orbits. \citet{2015ApJ...801..105X} found that the observed stellar ring structures in the disk are associated with radial waves with peaks above and below the Galactic plane. \citet{2015ApJ...801..105X} find an upward peak or overdensity of stars at 10 kpc, a downward peak at 12-14 kpc, then an upward peak at 16-18 kpc, and a downward peak at 20-24 kpc. These peaks are coincidentally near the positions of the predicted caustic rings at 10, 13, and 20 kpc.

\section{Summary and Conclusion}

In this paper we have presented the first study of the caustic ring model of the Milky Way halo \citep{2008PhRvD..78f3508D}, or ``Sikive halo", in relation to astronomical observations. We show that the caustic ring model is consistent with observations of stars in the Milky Way galaxy.

We presented the formalism for calculating the gravitational acceleration of a caustic ring halo. Our open source code for a caustic ring halo is available in the NEMO Stellar Dynamics Toolbox and the Milkyway@home client repository. We use this code to show that the caustic ring halo reproduces a roughly logarithmic halo, with large perturbations near the rings. The caustic ring halo has very little mass within 2 kpc of the Galactic center, resulting in a gravitational acceleration that points away from the Galactic center and towards the inner-most rings for distances less than 2 kpc. This outward acceleration may not be realized in practice, due to hierarchical merging and the presence of baryons.

We show that the caustic ring halo can reasonably match the known Galactic rotation curve. We were not able to confirm or rule out an association between the positions of the caustic rings and oscillations in the observed rotation curve, due to insufficient rotation curve data. The parameters for a Galactic model that includes a Hernquist bulge, Miyamoto-Nagai disk, and caustic halo are listed in Table \ref{rcfit}. The values for the disk mass, scale length, and scale height; and bulge mass and scale length that we found are comparable to those found in other studies which fit Galaxy component models to observations \citep{1998MNRAS.294..429D, 2005ApJ...631..838W, 2011MNRAS.414.2446M, 2014ApJ...794...59K}. 

We explore the effects of dark matter caustic rings on dwarf galaxy tidal disruption with N-body simulations. Simulations of the Sagittarius dwarf galaxy in a caustic ring halo potential with disk and bulge parameters that are tuned to match the Galactic rotation curve, match observations of the Sgr trailing tidal tails as far as 90 kpc from the Galactic center. Like the NFW halo, they are however unable to match the leading tidal tail. None of the caustic, NFW, or triaxial logarithmic halos are able to simultaneously match observations of the leading and trailing arms of the Sagittarius stream.

We further show that simulations of dwarf galaxies that move through caustic rings are qualitatively similar to those moving in a logarithmic halo. The dwarf galaxy disrupts more or less slowly depending on whether it moves on a planar or polar orbit in the caustic and logarithmic halo potentials. In simulations of dwarf galaxies that move just inside or outside of a caustic ring, the tidal debris is shown to be disrupted mostly to the interior or exterior of the ring respectively.

\acknowledgements

We gratefully acknowledge Pierre Sikivie for helpful discussions and clarifications about the caustic ring model. We acknowledge Peter Teuben for assisting with the integration of our code into the NEMO Stellar Dynamics Toolbox. We thank Heywood Tam for sending us his program used to calculate the gravitational field near a caustic ring. We also thank Larry Widrow for insightful discussions.

J.D. acknowledges support from NSF grant AST 10-09670, the NASA-NY Space Grant, the American Fellowship from AAUW, and generous gifts from the Marvin Clan, Babette Josephs, Manit Limlamai, and the MilkyWay@home volunteers.

\bibliographystyle{apj}

\end{document}